\documentclass[12pt]{article}
\usepackage{amssymb}
\usepackage{amsfonts}
\usepackage{youngtab}
\usepackage{amsmath,amssymb,latexsym,cite}
\usepackage{amsthm}
\usepackage{cite}
\usepackage[a-1b]{pdfx}    
\usepackage[]{hyperref}
\input xy
\xyoption{all}

\numberwithin{equation}{section}

\begin{document}

\vspace*{0.5in}

\begin{center}

{\large\bf Quantum K theory of symplectic Grassmannians}

\vspace{0.2in}

Wei Gu$^1$, 
Leonardo Mihalcea$^2$, Eric Sharpe$^3$, Hao Zou$^3$

\begin{tabular}{cc}
{\begin{tabular}{l}
$^1$ Center for Mathematical Sciences\\
Harvard University\\
Cambridge, MA  02138
\end{tabular}} &
{\begin{tabular}{l}
$^3$ Dep't of Physics\\
Virginia Tech\\
850 West Campus Dr.\\
Blacksburg, VA  24061
\end{tabular}}
\\
{\begin{tabular}{l}
$^2$ Dep't of Mathematics\\
Virginia Tech\\
225 Stanger St.\\
Blacksburg, VA  24061
\end{tabular}}
&
\end{tabular}

{\tt weigu@cmsa.fas.harvard.edu},
{\tt lmihalce@math.vt.edu},
{\tt ersharpe@vt.edu},
{\tt hzou@vt.edu}

$\,$

\end{center}

In this paper we discuss physical derivations of the quantum K theory
rings of symplectic Grassmannians.  We compare to standard presentations
in terms of Schubert cycles, but most of our work revolves around a proposed
description
in terms of two other bases, involving shifted Wilson lines and
$\lambda_y$ classes, which are motivated by and amenable to physics,
and which we also provide for ordinary Grassmannians.

\begin{flushleft}
August 2020
\end{flushleft}

\newpage

\tableofcontents

\newpage

\section{Introduction}

In physics folklore, going from $d$ dimensions to $d+1$ dimensions can often
be interpreted as some sort of K-theoretic uplift.  For example,
Donaldson/Seiberg-Witten theory in twisted four-dimensional
$N=2$ supersymmetric theories becomes a K-theoretic analogue
\cite{Gottsche:2006bm} in a five-dimensional $N=1$ supersymmetric theory.

Another such uplift has recently attracted attention:
quantum K theory \cite{g1,Lee:2001mb,g11,buch-mih1,rz,chap-perr,bcmp,gt} arises in descriptions of
three-dimensional supersymmetric gauge theories, `uplifting' 
Gromov-Witten invariants of two-dimensional (2,2) supersymmetric theories.
In particular, physics computations of
the quantum K theory invariants, analogues of
Gromov-Witten invariants, have been discussed in
\cite{Jockers:2018sfl,Jockers:2019wjh,Jockers:2019lwe,Nekrasov:2009uh,Gaiotto:2013bwa,Bullimore:2014awa} 
(see also \cite{Koroteev:2017nab,Ueda:2019qhg,Closset:2019hyt,Bullimore:2018jlp,Bullimore:2019qnt,Bullimore:2020nhv,Closset:2016arn,Closset:2017zgf,Aganagic:2017tvx,Yoshida:2014ssa,Smirnov:2020lhm,Aganagic:2017gsx,Wu:2020nis,Drukker:2019bev}).

The three-dimensional gauge theories in question are three-dimensional
gauged linear sigma models, and they have a few complications relative
to their two-dimensional counterparts.  The most important is the existence
of possible Chern-Simons terms.  In order to match the ordinary
quantum K theory arising in mathematics, one must, for example, pick
Chern-Simons levels carefully.  (Other values of the levels may
correspond to twisted quantum K theories described in \cite{rz}.)
A precise dictionary for such choices has been worked out for
three-dimensional $N=2$ supersymmetries theories without a superpotential;
however, as we shall see in this paper, that dictionary breaks down
in theories with a superpotential, and one of our results will be
a proposed extension to some examples with a superpotential.

Now, computing quantum K theory rings from GLSMs, much as with
quantum cohomology rings \cite{Morrison:1994fr}, relies on being able to work
over a nontrivial Coulomb branch, which is often complicated in theories
with a superpotential.  However, there do exist some clean examples of
theories with superpotential in which it is known that one can compute
quantum cohomology rings.  Low-degree hypersurfaces in projective spaces
are one set of examples, and another are symplectic Grassmannians,
as discussed in \cite{Gu:2020oeb,ot}.  In this paper, using the Chern-Simons
level ansatz mentioned above, we will compute quantum K theory rings
arising from physics for low-degree hypersurfaces in projective spaces
and Lagrangian Grassmannians, and compare to known mathematics results.

One point that may be of interest for mathematicians is that in these
cases, the quantum K theory relations arise as derivatives of a 
universal function, known in physics as the twisted one-loop effective
superpotential.  We will discuss this in detail later.

Another point that may be of interest for mathematicians is that
we provide two new descriptions of the quantum K theory rings of
ordinary and symplectic Grassmannians, in bases of shifted
Wilson lines and $\lambda_y$ classes, both of which are motivated by
physics.  We plan to address the mathematical details in
\cite{leom}.

We begin in section~\ref{sect:basics-overview} 
with a review of basics, including the quantum
K theory ring of ordinary Grassmannians, in order to make
this paper self-contained.  In section~\ref{sect:gkn:shifted} we describe the
quantum K theory ring of Grassmannians in a basis of shifted Wilson lines,
as motivated by physical considerations, and
anticipating a rigorous presentation in \cite{leom}.  
In section~\ref{sect:lambda:gkn} 
we describe the quantum K theory ring of Grassmannians
in another basis, of $\lambda_y$ classes of universal subbundles and
quotient bundles on the Grassmannian, also motivated by physics,
which we intend to describe rigorously in \cite{leom}.

In section~\ref{sect:levels} we briefly review the physical derivation
of quantum K theory rings for hypersurfaces in projective spaces,
as a warm-up exercise for symplectic Grassmannians.

In section~\ref{sect:sg} we finally turn to quantum K theory rings
of symplectic Grassmannians.  After reviewing basics of 
GLSM descriptions of symplectic Grassmannians, and checking that
physics correctly predicts known mathematics results for the case of
$LG(2,4)$, we give predictions for quantum K theory rings of symplectic
Grassmannians in a shifted Wilson line basis, and $\lambda_y$ class relations
for Lagrangian Grassmannians, which we intend to address
mathematically 
in \cite{leom}.
In section~\ref{sect:sgk2n:exs} 
we check these descriptions of the quantum K theory
ring in the cases of $LG(2,4)$ and $LG(3,6)$.
Finally, in appendix~\ref{app:tables} we list some results for
the quantum K theory ring of $LG(3,6)$.

Much of this paper focuses on Lagrangian and symplectic Grassmannians.
GLSMs for orthogonal Grassmannians were also discussed in
\cite{Gu:2020oeb,ot}; however, many of those GLSMs do not have nontrivial
Coulomb branches, and others have mixed Higgs/Coulomb branches, making
it impossible to apply the analysis of this paper.

\section{Review and ordinary Grassmannians}
\label{sect:basics-overview}

\subsection{Basics}  \label{sect:rev}

Briefly, quantum K theory arises in physics from $N=2$ supersymmetric
gauged linear sigma
models in three dimensions.  Consider a GLSM on a three-manifold of
the form $S^1 \times \Sigma$, for $\Sigma$ a Riemann surface.
This theory admits half-BPS Wilson lines, that are independent of 
motions along $\Sigma$.  Briefly, quantum K theory arises as the 
OPE algebra\footnote{
In passing, 
although a full topological twist of the three-dimensional theory does
not exist, one can (partially) twist, along $\Sigma$.
See \cite{Aganagic:2017tvx} for details.
}
of Wilson loops about the $S^1$.

Such OPE algebras can be computed by reduction to two dimensions.
One can build an effective two-dimensional (2,2) supersymmetric GLSM
with a Kaluza-Klein tower of fields.  Using zeta function regularization,
one can sum the contributions to the twisted one-loop effective action,
and obtain
\cite[equ'n (2.33)]{Closset:2016arn} 
\begin{eqnarray}
W(u,\nu) & = & \frac{1}{2} k^{ab} \left(\ln x_a \right) \left( \ln x_b \right)
 \: + \:
\frac{1}{2} k^{aF} \left( \ln x_a\right) \left( \ln y_F \right)
\nonumber \\
& &
\: + \:
\sum_a \left( \ln q_a \right) \left( \ln x_a \right)
\: + \: \sum_a \left( i \pi \sum_{\mu \: {\rm pos'}} \alpha^a_{\mu} \right)
 \left( \ln x_a \right)
\nonumber \\
& &
\: + \: 
\sum_i \left[ {\rm Li}_2\left( x^{\rho_i} y_i \right) 
\: + \: \frac{1}{4} \left( \rho_i(\ln x) + \ln y_i \right)^2 \right].
\label{eq:one-loop-w}
\end{eqnarray}
In the expression above, 
$i$ indexes fields, $y_i = \exp(2 \pi i \nu_i)$ encodes flavor symmetries,
$u_a = R \sigma_a$ for $R$ the radius of the three-dimensional $S^1$
and $\sigma_a$ the eigenvalues of the diagonalized adjoint-valued $\sigma$ in
the two-dimensional vector multiplet,
$x_a = \exp(2 \pi i u_a)$, and
\begin{equation}
x^{\rho_i} \: \equiv  \: \prod_a x_a^{\rho^a_i} \: = \: 
\exp\left( 2 \pi i \rho_i (u) \right).
\end{equation}
Because of their three-dimensional origins, the $\sigma$'s are periodic,
with periodicity which we will take to be
\begin{equation}
\sigma_a \: \sim \: \sigma_a + 1/R,
\end{equation}
under which the $x_a = \exp(2 \pi i R \sigma_a)$ are invariant.
(In passing, note that in the limit $R \rightarrow 0$, this becomes the
ordinary non-periodic scalar of a two-dimensional supersymmetric
theory.)
Also, in principle the $(\ln q)(\ln x)$ term could be recast
as a $(\ln x)(\ln y)$ term, but we have kept it separate for clarity.

The $\alpha^a_{\mu}$ on the second line are root vectors, contributing
a phase to $q_a$ (much as in e.g. \cite{Gu:2018fpm} and references therein).
The resulting phase could be absorbed into a shift of $q$, but we have
chosen to work in conventions in which they are kept explicit, so as to
match conventions of other sources.  For the case of gauge group
$U(k)$, it can be shown that
\begin{equation}
i \pi \sum_{\mu \: {\rm pos'}} \alpha^a_{\mu}
\: = \: i \pi (k-1)
\end{equation}
for all $a$,
so the effect will be to multiply $q$ by the phase $(-)^{k-1}$.

For later use, a handy identity is
\begin{equation}
x \frac{\partial}{\partial x} {\rm Li}_2(x) \: = \:
{\rm Li}_1(x) \: = \: - \ln(1 - x).
\end{equation}

For three-dimensional $N=2$ theories without a superpotential,
the
Chern-Simons levels describing the ordinary quantum K theory rings
are determined by starting with an $N=4$ theory
in three dimensions
and integrating out fields to build the given $N=2$ theory.
For an abelian gauge theory in which $Q^i_a$ denotes the charge of the
$i$th chiral superfield under the $a$th $U(1)$ factor,
\begin{equation}
k^{ab} \: = \: - \frac{1}{2} \sum_i Q^i_a Q^i_b.
\end{equation}
(In a $U(1)$ gauge theory, this coincides with 
$U(1)_{-1/2}$ quantization \cite[section 2.2]{Closset:2019hyt}.)

Mixed gauge-flavor levels are determined in the same fashion.
If the $i$th chiral superfield has R-charge $r_i$,
\begin{equation}
k^{aR} \: = \: - \frac{1}{2} \sum_i Q^i_a (r_i - 1).
\end{equation}
(See also \cite[section 2.2]{Jockers:2019lwe} 
for a discussion of windows of levels for
which one recovers quantum K theory.)
For example, for a nonabelian simple gauge group $G$ 
with matter in representation $R$,
there is a contribution to the pertinent level $k_G$ from matter in
representation $R$ given by
\begin{equation}
 - (1/2) T_2(R),
\end{equation}
where $T_2(R)$ is the quadratic index of the representation $R$,
normalized so that, in $SU(k)$, $T_2({\rm fundamental}) = 1$.

The expression for the twisted one-loop effective superpotential
has been written in terms of an abelianization of the gauge group.
Let us now consider a nonabelian gauge theory.  Specifically, consider
a $U(k)$ gauge theory.
Here, there are two levels, one for the overall trace $U(1)$,
another for $SU(k)$.  Using the facts that
\begin{eqnarray}
{\rm tr}_{U(k)} \sigma^2 & = & \sum_a \sigma_a^2, 
\\
{\rm tr}_{U(1)} \sigma^2 & = & \frac{1}{k} \left( \sum_a \sigma_a \right)^2,
\\
{\rm tr}_{SU(k)} \sigma^2 & = & {\rm tr}_{U(k)} \sigma^2 \: - \:
{\rm tr}_{U(1)} \sigma^2,
\end{eqnarray}
where $\sigma$ is adjoint-valued and the $\sigma_a$ its eigenvalues,
we see that $k_{U(1)}$ couples to
\begin{equation}
\frac{1}{k} \left( \sum_a \ln x_a \right)^2,
\end{equation}
and $k_{SU(k)}$ couples to
\begin{equation}
\sum_a (\ln x_a)^2 \: - \: \frac{1}{k} \left( \sum_a \ln x_a \right)^2.
\end{equation}
As a result, we take
\begin{eqnarray}
k^{ab} u_a u_b & = &
k_{SU(k)} \left[ \sum u_a^2 - \frac{1}{k} \left( \sum_a u_a \right)^2 \right]
\: + \:
\frac{ k_{U(1)} }{k} \left( \sum_a u_a \right)^2,
\\
& = &
k_{SU(k)} \sum_a u_a^2 \: + \:
\frac{ k_{U(1)} - k_{SU(k)} }{k} \left( \sum_a u_a \right)^2.
\end{eqnarray}

Furthermore, we take
\begin{eqnarray}
\frac{1}{4} \sum_i \left( \rho_i (\ln x) \right)^2 
& = & \frac{1}{4} \sum_i \sum_a \left[ \sum_b \rho^b_{i a} \ln x_b \right]^2.
\end{eqnarray}
For example, for $n$ copies of the fundamental representation of $U(k)$,
\begin{equation}
\rho^b_{ia} \: = \: \delta^b_a
\end{equation}
(independent of the flavor index $i$), and so
\begin{equation}
\frac{1}{4} \sum_i \left( \rho_i (\ln x) \right)^2 
\: = \:
\frac{n}{4} \sum_a \left( \ln x_a \right)^2
\end{equation}
in this case.

So far, we have discussed the Chern-Simons levels that one should pick
in a three-dimensional $N=2$ supersymmetric theory without superpotential,
so as to reproduce ordinary quantum K theory (as we will verify in examples
shortly).  In principle, there is one other matter about which we should
also be careful, namely the `topological vacua' 
\cite{Intriligator:2013lca,Bullimore:2019qnt,Bullimore:2020nhv}.
These are closely related to discrete Coulomb vacua in
two-dimensional GLSMs \cite{Melnikov:2005hq,Melnikov:2006kb}.
If they are present in a given GLSM phase, then they should be considered
part of the geometry of that phase, a modification of the target space,
which would complicate efforts to derive quantum K theory relations from
physics.  As a result, to hope for a derivation of quantum K theory,
one must require that there are no topological vacua in that phase,
which typically constrains possible Chern-Simons levels.

Next, let us consider the operators.
In the reduction to two dimensions,
the Wilson line operators become two-dimensional operators of the
form
\begin{equation}
{\rm Tr}\, \exp \left( 2 \pi i R \sigma \right),
\end{equation}
where $\sigma$ denotes the adjoint-valued scalar in the
two-dimensional (2,2) supersymmetric vector multiplet,
and then derive OPE relations from the equations of motion for $\sigma$
derived from the twisted-one-loop effective action,
in the same fashion as one ordinarily
derives quantum cohomology relations in two-dimensional theories
\cite{Morrison:1994fr}.

Mathematically, those Wilson line operators correspond in K theory to
locally-free sheaves.  The quantum K theory relations are typically
stated as relations between varieties -- for Grassmannians,
Schubert varieties.  

For later use, in two-dimensional nonabelian theories,
it is important to take into account the excluded loci when deriving
quantum cohomology relations.  For example, in a $U(k)$ gauge
theory, on the Coulomb branch, $\sigma_a \neq \sigma_b$ for
$a \neq b$.  The $x$ fields obey analogous relations,
in this case $x_a \neq x_b$ for $a \neq b$.  This is not a stronger
constraint, because the $\sigma_a$ descending from three dimensions
are cylinder-valued, and so $\sigma_a \neq \sigma_b$ if and only if
$x_a \neq x_b$.

If $R$ denotes the radius of the three-dimensional
$S^1$, then the ordinary quantum cohomology relations are obtained
as the $R \rightarrow 0$ limit of the relations amongst the Schubert
varieties.  (In fact, in this limit, the operators corresponding to the
Schubert varieties reduce to Schur polynomials in the $\sigma$'s, which
are precisely the operators describing quantum cohomology rings in GLSMs.)
In such limits, it is important to distinguish the three-dimensional
$q$ (henceforward $q_{3d}$) from the ordinary two-dimensional $q$ ($q_{2d}$)
arising in quantum cohomology computations in GLSMs.
In three dimensions, $q$ is dimensionless, but the two-dimensional
version is not.  Instead, it is related by dimensional-transmutation,
giving a factor
$\Lambda^{b_0}$, where $b_0$ is determined by the two-dimensional
beta function (axial R-symmetry anomaly).
For example, for a $U(k)$
theory with $n$ fundamentals, $b_0 = n$.
Absorbing dimensions into $R$, in that case we have
\begin{equation}   \label{eq:3d2d:funds}
q_{3d} \: = \: R^n q_{2d}.
\end{equation}

The ordinary quantum cohomology ring admits a grading,
which in the two-dimensional theory corresponds to the axial R-symmetry
or BRST grading.  In three dimensions, there is no such symmetry,
and indeed, the quantum K theory products are not consistent with
such a symmetry.  We shall see this in examples later in this paper.

\subsection{Review of projective spaces}
\label{sect:rev:pn}

For projective spaces, the quantum K theory ring is identical to the
quantum cohomology ring.  In this section, we will establish that fact
in terms of the twisted one-loop effective superpotential and its
dilogarithms for the three-dimensional theory.

In the physical realization of the quantum cohomology ring,
we identify cohomology classes with Young tableaux, which are
identified with Schur polynomials.  For projective spaces, this
dictionary takes the following form:
\begin{eqnarray}
\tiny\yng(1) & = & \sigma, 
\\
\tiny\yng(2) & = & \sigma^2,
\\
\tiny\yng(3) & = & \sigma^3,
\end{eqnarray}
and so forth.  On ${\mathbb P}^n$, there is the relation
$\sigma^{n+1} \sim q$.

In the physical realization of quantum K theory,
we identify K theory classes with Young tableaux, which
are identified with Chern characters $\exp(2 \pi i \sigma)$ as discussed
in section~\ref{sect:rev}.

In the case of a GLSM for the projective space ${\mathbb P}^n$,
the two-dimensional twisted one-loop effective superpotential
derived from equation~(\ref{eq:one-loop-w}) is
\begin{equation}
W \: = \: \frac{1}{2} \left(k + \frac{n+1}{2} \right) \left( \ln x \right)^2
\: + \: \left( \ln q \right) \left( \ln x \right)
\: + \: \sum_{i=1}^{n+1} {\rm Li}_2(x).
\end{equation}
(The $(n+1)/4 (\ln x)^2$ term arises from the $(1/4) \rho(\ln x)^2$ term
in equation~(\ref{eq:one-loop-w}).)  In
$U(1)_{-1/2}$ quantization in this theory,
\begin{equation}
k \: = \: - \frac{n+1}{2},
\end{equation}
so we see that the first term in the superpotential $W$ drops out, leaving
\begin{equation}
W \: = \:  \left( \ln q \right) \left( \ln x \right)
\: + \: \sum_{i=1}^{n+1} {\rm Li}_2(x).
\end{equation}

From this one derives the equations of motion
\begin{equation}  \label{eq:pn:qk}
(1 - x)^{n+1} \: = \: q.
\end{equation}
In terms of K theory, we identify the operator $W_{\tiny\yng(1)} = x$ with 
$S = {\cal O}(-1)$ in the Grothendieck group of ${\mathbb P}^n$,
and so we have the relation
\begin{equation}
\left( 1 - S \right)^{n+1} \: = \: q.
\end{equation}
In terms of Schubert varieties, the hyperplane class ${\cal O}_{\tiny\yng(1)}$
is the cokernel of an inclusion $S \hookrightarrow {\cal O}$:
\begin{equation}
0 \: \longrightarrow \: S \: \longrightarrow \: {\cal O} \:
\longrightarrow \: {\cal O}_{\tiny\yng(1)} \: \longrightarrow \: 0,
\end{equation}
hence in terms of K theory,
\begin{equation}
S \: + \: {\cal O}_{\tiny\yng(1)} \: = \: 1,
\end{equation}
or simply $W_{\tiny\yng(1)} = 1 - {\cal O}_{\tiny\yng(1)}$,
so we have the quantum K theory relation
\begin{equation} \label{eq:pn-qk}
\left(  {\cal O}_{\tiny\yng(1)} \right)^{n+1} \: = \: q.
\end{equation}

Now, for completeness, let us consider the $R \rightarrow 0$ limit,
to recover ordinary quantum cohomology.
In this limit, the Schubert varieties reduce to Schur polynomials in
$\sigma$'s:
\begin{eqnarray}
{\cal O}_{\tiny\yng(1)} & = & 1 - W_{\tiny\yng(1)} \: = \: 
1 - \exp\left( 2 \pi i R \sigma \right),
\\
& \mapsto & 1 - \left( 1 + 2 \pi i R \sigma \right) \: = \:
- 2 \pi i R \sigma.
\end{eqnarray}
The quantum K theory relation~(\ref{eq:pn-qk}) becomes
\begin{equation}
(- 2 \pi i R)^{n+1} \sigma^{n+1} \: = \: q_{3d},
\end{equation}
and since from section~\ref{sect:rev} we know that
$q_{3d} = R^{n+1} q_{2d}$ in this case, we have
\begin{equation}
\sigma^{n+1} \: \propto \: q_{2d},
\end{equation}
as both sides are at the same (leading) order in the radius $R$.
This is, of course, the well-known quantum cohomology ring relation
for ${\mathbb P}^n$.

\subsection{Review of ordinary Grassmannians and Schubert class bases}

Physical realizations of quantum K theory rings for ordinary
Grassmannians have recently been discussed in detail in
\cite{Jockers:2019lwe,Ueda:2019qhg}, so our overview in this section
will be brief, focused on setting up some ideas for use in later sections.  
In particular, we include this discussion because in subsequent
sections we will give novel alternative descriptions of the quantum
K theory ring of Grassmannians (in terms of the shifted Wilson line basis
and $\lambda_y$ classes), which will themselves later be used to describe the
quantum K theory of Lagrangian Grassmannians and to compare to physics
predictions.

Briefly, a Grassmannian $G(k,n)$ is realized by a $U(k)$ gauge
theory with $n$ fundamentals.  The levels are as follows:
\begin{eqnarray}
k_{U(1)} & = & - n/2, 
\\
k_{SU(k)} & = & k - n/2
\end{eqnarray}
(see for example \cite[equ'n (2.5)]{Jockers:2019lwe},
\cite[equ'n (4.15)]{Ueda:2019qhg}).
(In $k_{SU(k)}$, for example, if we think of deriving the $N=2$ action from
a three-dimensional $N=4$ theory, the term $k$ arises from integrating out the
extra $N=2$ chiral multiplet needed to build the $N=4$ vector multiplet,
and the $-n/2$ from integrating out half of the $N=4$ hypermultiplets.)

Assembling the details from section~\ref{sect:rev}, we see that the
two-dimensional twisted one-loop effective superpotential is
\begin{eqnarray}
W & = &
\frac{1}{2} k_{SU(k)} \sum_a \left( \ln x_a \right)^2
\: + \:
\frac{ k_{U(1)} - k_{SU(k)} }{2k} \left( \sum_a \ln x_a \right)^2
\nonumber \\
& &
\: + \: \left(\ln (-)^{k-1} q\right) \sum_a \ln x_a \: + \:
n \sum_a {\rm Li}_2\left(  x_a \right) 
\: + \: \frac{n}{4} \sum_a \left( \ln x_a \right)^2,
\\
& = & \frac{k}{2} \sum_a \left( \ln x_a \right)^2
\: - \: \frac{1}{2} \left( \sum_a \ln x_a \right)^2
\nonumber \\
& & \: + \: \left(\ln (-)^{k-1} q\right) \sum_a \ln x_a \: + \:
n \sum_a {\rm Li}_2\left( x_a \right) .
\end{eqnarray}
The equations of motion are then
\begin{equation} \label{eq:gkn-eom}
(-)^{k-1} q x_a^k \: = \: (1-x_a)^n
\left( \prod_b x_b \right) .
\end{equation}
In this theory, we can interpret permutation-invariant (Schur) 
polynomials in the $x_a$
as either Wilson lines or, equivalently, as Schur functors in the
universal subbundle $S \rightarrow G(k,n)$, a perspective we will utilize
later.

A two-dimensional nonabelian gauge theory typically has an `excluded' locus on
its Coulomb branch (see e.g. \cite{Gu:2018fpm} for details).  For a $U(k)$
gauge theory, that excluded locus forbids vacua in which $\sigma_a =
\sigma_b$ for $a \neq b$, which means in the present case that we can
always take $x_a \neq x_b$, and algebraically cancel out factors of 
$x_a - x_b$.  In addition, since each $x$ is an exponential of $\sigma$,
we see that $x$ can never vanish for any finite value of $\sigma$,
so we can always take $x_a \neq 0$.

To make this concrete, and to set up later computations,
we will work through the details for
the example of the Grassmannian $G(2,4)$.
This is perhaps the simplest example in which
the quantum K theory ring differs from the quantum cohomology ring.
Both the quantum cohomology ring and the quantum K theory ring
have additive generators counted by Young tableaux fitting inside a $2 \times 2$
box, but they differ in their product structures.

First, we compute the quantum K theory of $G(2,4)$ determined by physics.
Here, each $W_R$ is given by a Schur polynomial (determined by the
representation $R$) in the $x_a$.  For example,
\begin{eqnarray}
W_{\tiny\yng(1)} & = & x_1 + x_2,
\\
W_{\tiny\yng(2)} & = & x_1^2 + x_2^2 + x_1 x_2,
\\
W_{\tiny\yng(1,1)} & = & x_1 x_2,
\\
W_{\tiny\yng(2,1)} & = & x_1^2 x_2 + x_1 x_2^2,
\\
W_{\tiny\yng(2,2)} & = & x_1^2 x_2^2,
\end{eqnarray}
where $x_a = \exp(2 \pi i R \sigma_a)$. 
The equations of motion~(\ref{eq:gkn-eom}) become
\begin{equation}   \label{eq:g24:eom1}
- q x_1 \: = \: (1 - x_1)^4 ( x_2), \: \: \:
- q x_2 \: = \: (1 - x_2)^4 (x_1 )
\end{equation}
(where we have used the fact that $x_a \neq 0$, since they are exponentials,
to cancel out common factors).
This implies
\begin{equation}  \label{eq:g24-eom-res1}
x_2^2 (1 - x_1)^4 \: = \: x_1^2 (1 - x_2)^4,
\end{equation}
and cancelling out a common factor of $x_1 - x_2$ (due to the excluded locus
condition),
the difference is
\begin{equation}  \label{eq:g24-eom-res2}
- q \: = \: -1 + 6 x_1 x_2 - 4 x_1 x_2 (x_1 + x_2) + x_1 x_2 (x_1^2 + x_1 x_2
+ x_2^2).
\end{equation}
Similarly,
\begin{equation} \label{eq:g24-eom-res3}
- 2 q x_1 x_2 \: = \: x_2^2 (1 - x_1)^4 + x_1^2 (1 - x_2)^4.
\end{equation}

After factoring out $x_1 - x_2$ (which can never vanish since $x_1 = x_2$
is on the excluded locus),
equation~(\ref{eq:g24-eom-res1}) becomes
\begin{equation}  \label{eq:g24-eom-res1a}
W_{\tiny\yng(1)} \cdot W_{\tiny\yng(2,2)} \: = \:
W_{\tiny\yng(1)} - 4 W_{\tiny\yng(1,1)} + 4 W_{\tiny\yng(2,2)}
\end{equation}
equation~(\ref{eq:g24-eom-res2}) becomes
\begin{equation}
- q \: = \: -1 + 6 W_{\tiny\yng(1,1)} - 4 W_{\tiny\yng(2,1)} + W_{\tiny\yng(1,1)}
\cdot W_{\tiny\yng(2)},
\end{equation}
and equation~(\ref{eq:g24-eom-res3}) becomes
\begin{equation}
W_{\tiny\yng(1,1)} \cdot W_{\tiny\yng(2,2)} \: = \:
q W_{\tiny\yng(1,1)} + W_{\tiny\yng(2)} - 4 W_{\tiny\yng(2,1)}
+ 6 W_{\tiny\yng(2,2)}.
\end{equation}

The OPEs for the Wilson line operators $W_T$ can now be derived
algebraically, using the relations above:
\begin{eqnarray}
W_{\tiny\yng(1)} \cdot W_{\tiny\yng(1)}
& = & W_{\tiny\yng(2)} + W_{\tiny\yng(1,1)},
\\
W_{\tiny\yng(1,1)} \cdot W_{\tiny\yng(1)} 
& = & W_{\tiny\yng(2,1)},
\\
W_{\tiny\yng(2)} \cdot W_{\tiny\yng(1)}
& = & 4(-q+1) -6 W_{\tiny\yng(1)} + 4 q W_{\tiny\yng(1,1)} + 4 W_{\tiny\yng(2)}
+ (-q+1) W_{\tiny\yng(2,1)},
\\
W_{\tiny\yng(2,1)} \cdot W_{\tiny\yng(1)}
& = & -q + 1 - 6 W_{\tiny\yng(1,1)} + 4 W_{\tiny\yng(2,1)} + W_{\tiny\yng(2,2)},
\\
W_{\tiny\yng(2,2)} \cdot W_{\tiny\yng(1)}
& = & W_{\tiny\yng(1)} - 4 W_{\tiny\yng(1,1)} + 4 W_{\tiny\yng(2,2)},
\\
W_{\tiny\yng(1,1)} \cdot W_{\tiny\yng(1,1)} 
& = & W_{\tiny\yng(2,2)},
\\
W_{\tiny\yng(2)} \cdot W_{\tiny\yng(1,1)}
& = & -q + 1 - 6 W_{\tiny\yng(1,1)} + 4 W_{\tiny\yng(2,1)},
\\
W_{\tiny\yng(2,1)} \cdot W_{\tiny\yng(1,1)}
& = & W_{\tiny\yng(1)} - 4 W_{\tiny\yng(1,1)} + 4 W_{\tiny\yng(2,2)}
\: = \: W_{\tiny\yng(2,2)} \cdot W_{\tiny\yng(1)},
\label{eq:g24:21-11}
\\
W_{\tiny\yng(2,2)} \cdot W_{\tiny\yng(1,1)} 
& = & 
q W_{\tiny\yng(1,1)} + W_{\tiny\yng(2)} - 4 W_{\tiny\yng(2,1)}
+ 6 W_{\tiny\yng(2,2)},
\\
W_{\tiny\yng(2)} \cdot W_{\tiny\yng(2)} 
& = & 
(-q+1)^2 + 15 (-q+1) + 4 (-q - 5) W_{\tiny\yng(1)} 
+ 10 W_{\tiny\yng(2)}
\nonumber \\
& & \hspace*{0.5in}
- (-22 q + 6) W_{\tiny\yng(1,1)} + 4 (-q+1) W_{\tiny\yng(2,1)}
+ (-q+1) W_{\tiny\yng(2,2)},
\\
W_{\tiny\yng(2,1)} \cdot W_{\tiny\yng(2)}
& = & 4 (-q+1) + (-q+1) W_{\tiny\yng(1)} - 24 W_{\tiny\yng(1,1)}
+ 10 W_{\tiny\yng(2,1)} + 4 W_{\tiny\yng(2,2)}, 
\\
W_{\tiny\yng(2,2)} \cdot W_{\tiny\yng(2)}
& = & (-q - 15) W_{\tiny\yng(1,1)} + 4 W_{\tiny\yng(1)} +
10 W_{\tiny\yng(2,2)},
\\
W_{\tiny\yng(2,1)} \cdot W_{\tiny\yng(2,1)}
& = &
4 W_{\tiny\yng(1)} + W_{\tiny\yng(2)} - 15 W_{\tiny\yng(1,1)} - 
4 W_{\tiny\yng(2,1)} + 16 W_{\tiny\yng(2,2)},
\\
W_{\tiny\yng(2,2)} \cdot W_{\tiny\yng(2,1)}
& = &
 4 q W_{\tiny\yng(1,1)} + 4 W_{\tiny\yng(2)} - 15 W_{\tiny\yng(2,1)}
+ 20 W_{\tiny\yng(2,2)},
\\
W_{\tiny\yng(2,2)} \cdot W_{\tiny\yng(2,2)}
& = &
-q + 1 - 4 W_{\tiny\yng(1)} + (10 + 6 q) W_{\tiny\yng(1,1)}
+ 6 W_{\tiny\yng(2)} - 20 W_{\tiny\yng(2,1)}
\nonumber \\
& & \hspace*{0.5in}
+ (20 + q) W_{\tiny\yng(2,2)}.
\end{eqnarray}

Now, let us compare to Schubert varieties.  Let ${\cal O}_T$ denote the
Schubert variety corresponding to a Young tableau $T$.
From \cite[chapter 1.5]{gh}, ${\cal O}_{\tiny\yng(1)}$ corresponds to
a hyperplane in the Pl\"ucker embedding, hence the codimension-one
Schubert variety ${\cal O}_{\tiny\yng(1)}$ is resolved by
\begin{equation}
0 \: \longrightarrow \: \det S \: \longrightarrow \: {\cal O} \:
\longrightarrow \: {\cal O}_{\tiny\yng(1)} \: \longrightarrow \: 0.
\end{equation}
Taking Chern characters, this implies
\begin{equation}
{\rm ch}\left( {\cal O} \right) \: = \: {\rm ch}\left( \det S \right)
\: + \: {\rm ch}\left( {\cal O}_{\tiny\yng(1)} \right).
\end{equation}
Now, ch$({\cal O}) = 1$, and for $G(2,n)$, ch$(\det S) = W_{\tiny\yng(1,1)}$,
hence we have
\begin{equation}
1 \: = \: W_{\tiny\yng(1,1)} \: + \: {\rm ch}\left( {\cal O}_{\tiny\yng(1)}
\right).
\end{equation}
Other cases are analogous, but considerably more complicated in general.
We simply list the result below:
\begin{eqnarray}
{\cal O}_{\tiny\yng(1)} & = & 1 - W_{\tiny\yng(1,1)},  \label{eq:g24:o1}
\\
{\cal O}_{\tiny\yng(1,1)} & = & 1 - W_{\tiny\yng(1)} +
W_{\tiny\yng(1,1)},
\\
{\cal O}_{\tiny\yng(2)} & = & 1 - 3 W_{\tiny\yng(1,1)} +
W_{\tiny\yng(2,1)},  \label{eq:g24:o2}
\\
{\cal O}_{\tiny\yng(2,1)} & = & 1 - W_{\tiny\yng(1)}
+ W_{\tiny\yng(2,1)} - W_{\tiny\yng(2,2)},
\\
{\cal O}_{\tiny\yng(2,2)} & = & 1 - 2 W_{\tiny\yng(1)}
+ W_{\tiny\yng(2)} + 3 W_{\tiny\yng(1,1)} - 2 W_{\tiny\yng(2,1)}
+ W_{\tiny\yng(2,2)}.
\end{eqnarray}

After changing basis to the Schubert varieties above,
it is straightforward to show that
\begin{eqnarray}
{\cal O}_{\tiny\yng(1)} \cdot {\cal O}_{\tiny\yng(1)} & = &
{\cal O}_{\tiny\yng(1,1)} + {\cal O}_{\tiny\yng(2)} - {\cal O}_{\tiny\yng(2,1)},
\\
{\cal O}_{\tiny\yng(1,1)} \cdot {\cal O}_{\tiny\yng(1)} & = &
{\cal O}_{\tiny\yng(2,1)},
\\
{\cal O}_{\tiny\yng(2)} \cdot {\cal O}_{\tiny\yng(1)} & = &
{\cal O}_{\tiny\yng(2,1)},
\\
{\cal O}_{\tiny\yng(2,1)} \cdot {\cal O}_{\tiny\yng(1)} & = &
{\cal O}_{\tiny\yng(2,2)} + q - q {\cal O}_{\tiny\yng(1)},
\\
{\cal O}_{\tiny\yng(2,2)} \cdot {\cal O}_{\tiny\yng(1)} & = &
 q {\cal O}_{\tiny\yng(1)},
\\
{\cal O}_{\tiny\yng(1,1)} \cdot {\cal O}_{\tiny\yng(1,1)} & = &
{\cal O}_{\tiny\yng(2,2)},
\\
{\cal O}_{\tiny\yng(2)} \cdot {\cal O}_{\tiny\yng(1,1)} & = &  q,
\\
{\cal O}_{\tiny\yng(2,1)} \cdot {\cal O}_{\tiny\yng(1,1)} & = &
 q {\cal O}_{\tiny\yng(1)},
\\
{\cal O}_{\tiny\yng(2,2)} \cdot {\cal O}_{\tiny\yng(1,1)} & = &
 q {\cal O}_{\tiny\yng(2)},
\\
{\cal O}_{\tiny\yng(2)} \cdot {\cal O}_{\tiny\yng(2)} & = &
{\cal O}_{\tiny\yng(2,2)},
\\
{\cal O}_{\tiny\yng(2,1)} \cdot {\cal O}_{\tiny\yng(2)} & = &
 q {\cal O}_{\tiny\yng(1)},
\\
{\cal O}_{\tiny\yng(2,2)} \cdot {\cal O}_{\tiny\yng(2)} & = &
 q {\cal O}_{\tiny\yng(1,1)},
\\
{\cal O}_{\tiny\yng(2,1)} \cdot {\cal O}_{\tiny\yng(2,1)} & = &
 q {\cal O}_{\tiny\yng(1,1)} + q {\cal O}_{\tiny\yng(2)} -
q {\cal O}_{\tiny\yng(2,1)},
\\
{\cal O}_{\tiny\yng(2,2)} \cdot {\cal O}_{\tiny\yng(2,1)} & = &
 q {\cal O}_{\tiny\yng(2,1)},
\\
{\cal O}_{\tiny\yng(2,2)} \cdot {\cal O}_{\tiny\yng(2,2)} & = & q^2.
\end{eqnarray}
These match
the quantum K theory relations known to mathematics,
see e.g. \cite[example 5.9]{buch-mih1},
and have also been previously derived in physics, see e.g.
\cite[table (4.8)]{Jockers:2019lwe},
\cite[table (4.30)]{Ueda:2019qhg}.

Now, let us compare the $R \rightarrow 0$ limit
to quantum cohomology relations.
The quantum cohomology of $G(2,4)$ is
described by Schur polynomials
\begin{eqnarray}
\tiny\yng(1) & = & \sigma_1 + \sigma_2, 
\\
\tiny\yng(2) & = & \sigma_1^2 + \sigma_2^2 + \sigma_1 \sigma_2,
\\
\tiny\yng(1,1) & = & \sigma_1 \sigma_2,
\\
\tiny\yng(2,1) & = & \sigma_1^2 \sigma_2 + \sigma_1 \sigma_2^2,
\\
\tiny\yng(2,2) & = & \sigma_1^2 \sigma_2^2.
\end{eqnarray}
Each of these generators cohomology of degree determined by the total
number of boxes.

From the twisted one-loop effective superpotential, we have
\begin{equation}
\sigma_a^4 \: = \: -q,
\end{equation}
and the excluded locus condition is $\sigma_1 \neq \sigma_2$.
Since
\begin{equation}
\sigma_1^4 - \sigma_2^4 \: = \:
(\sigma_1 - \sigma_2) \left( \sigma_1^3 + \sigma_1^2 \sigma_2 +
\sigma_1 \sigma_2^2 + \sigma_2^3 \right),
\end{equation}
and we know $\sigma_1 \neq \sigma_2$, we therefore derive the condition
\begin{equation}
 \sigma_1^3 + \sigma_1^2 \sigma_2 +
\sigma_1 \sigma_2^2 + \sigma_2^3 \: = \: 0.
\end{equation}

By algebraically manipulating the equations above, we find that
the relations defining quantum cohomology of $G(2,4)$ include
\begin{eqnarray}
(\tiny\yng(1))^2 &  = & \tiny\yng(2) + \tiny\yng(1,1),
\label{eq:g24:qc:1-1}
\\
\tiny\yng(1) \cdot \tiny\yng(2) & = & \tiny\yng(2,1),
\\
\tiny\yng(1) \cdot \tiny\yng(1,1) & = & \tiny\yng(2,1),
\\
\tiny\yng(1) \cdot \tiny\yng(2,1) & = & \tiny\yng(2,2) + q,
\label{eq:g24:qc:1-21}
\\
\tiny\yng(1) \cdot \tiny\yng(2,2) & = &  q \, \tiny\yng(1),
\\
\left( \tiny\yng(1,1) \right)^2 & = & \tiny\yng(2,2),
\\
\tiny\yng(1,1) \cdot \tiny\yng(2) & = & q,
\\
\tiny\yng(1,1) \cdot \tiny\yng(2,1) & = &  q \, \tiny\yng(1),
\\
\tiny\yng(1,1) \cdot \tiny\yng(2,2) & = &  q \, \tiny\yng(2),
\\
\left(\tiny\yng(2) \right)^2 & = & \tiny\yng(2,2),
\\
\tiny\yng(2) \cdot \tiny\yng(2,1) & = &  q \, \tiny\yng(1),
\\
\tiny\yng(2) \cdot \tiny\yng(2,2) & = &  q \, \tiny\yng(1,1),
\\
\left( \tiny\yng(2,1) \right)^2 & = &  q \left( \tiny\yng(2) + \tiny\yng(1,1)
\right),
\\
\tiny\yng(2,1) \cdot \tiny\yng(2,2) & = &
q \, \tiny\yng(2,1),
\\
\left( \tiny\yng(2,2) \right)^2 & = & q^2.
\end{eqnarray}

The Schur polynomials above arise as the $R \rightarrow 0$ limit of the
operators ${\cal O}_T$ corresponding to Schubert varieties
(up to irrelevant factors).
For example,
\begin{eqnarray}
{\cal O}_{\tiny\yng(1)} & = & 1 - W_{\tiny\yng(1,1)} \: = \:
1 - \exp(2 \pi iR(\sigma_1 + \sigma_2)),
\\
& \mapsto & 1 - (1 + (2\pi i)R \sigma_1 + (2\pi i) R \sigma_2) \: = \: - (2\pi i)R \sigma_1 - (2\pi i)R \sigma_2
\\
& & = \: - (2\pi i)R \, {\tiny\yng(1)},
\end{eqnarray}
and similarly,
\begin{eqnarray}
{\cal O}_{\tiny\yng(1,1)} & \mapsto & R^2 \sigma_1 \sigma_2 \: = \:
R^2 \, {\tiny\yng(1,1)},
\\
{\cal O}_{\tiny\yng(2)} & \mapsto & R^2 \left( \sigma_1^2 + \sigma_2^2 +
\sigma_1 \sigma_2 \right) \: = \:
R^2 \, {\tiny\yng(2)},
\\
{\cal O}_{\tiny\yng(2,1)} & \mapsto & - R^3 \sigma_1^2 \sigma_2 - 
R^3 \sigma_1 \sigma_2^2 \: = \: - R^3 \, {\tiny\yng(2,1)},
\\
{\cal O}_{\tiny\yng(2,2)} & \mapsto & + R^4 \sigma_1^2 \sigma_2^2
\: = \: R^4 \, {\tiny\yng(2,2)},
\end{eqnarray}
where for simplicity we have suppressed factors such as $2 \pi i$
multiplying
$\sigma$'s.

If we take the limit $R \rightarrow 0$ of the
products ${\cal O}_T {\cal O}_{T'}$ above, then we can recover the
products of ordinary quantum cohomology; however, some subtleties
regarding the difference between scalings of $q$ in three dimensions
and two dimensions must be taken into account, as discussed
in section~\ref{sect:rev}.  Specifically, for $G(2,4)$, since there
are four fundamentals, equation~(\ref{eq:3d2d:funds}) becomes
\begin{equation}
q_{3d} \: = \: R^4 q_{2d}.
\end{equation}
This becomes important when determining which terms survive the
$R \rightarrow 0$ limit.

For example, in the multiplication tables above,
the quantum K theory relation
\begin{equation}
{\cal O}_{\tiny\yng(2,1)} \cdot {\cal O}_{\tiny\yng(1)} \: = \:
{\cal O}_{\tiny\yng(2,2)} \: + \: q_{3d} \: - \: q_{3d} {\cal O}_{\tiny\yng(1)}
\end{equation}
becomes
\begin{equation}
R^4 \left( {\tiny\yng(2,1)} \cdot {\tiny\yng(1)} \right) \: = \:
R^4 \, {\tiny\yng(2,2)} \: + \: R^4 q_{2d} \: - \: R^5 q_{2d} {\tiny\yng(1)}
\end{equation}
for small $R$, where again we have suppressed irrelevant factors.
In the limit $R \rightarrow 0$, the last term, proportional to $R^5$,
is subleading, and we are left with the quantum cohomology
relation~(\ref{eq:g24:qc:1-21}), as expected.
For another example, for small $R$, the quantum K theory relation
\begin{equation}
{\cal O}_{\tiny\yng(1)} \cdot {\cal O}_{\tiny\yng(1)} \: = \:
{\cal O}_{\tiny\yng(1,1)} \: + \:
{\cal O}_{\tiny\yng(2)} \: - \:
{\cal O}_{\tiny\yng(2,1)}
\end{equation}
becomes
\begin{equation}
R^2 \left( {\tiny\yng(1)} \cdot {\tiny\yng(1)} \right) \: = \:
R^2 \left( {\tiny\yng(1,1)} + {\tiny\yng(2)} \right)
\: - \:
R^3 \, {\tiny\yng(2,1)}.
\end{equation}
In the $R \rightarrow 0$ limit, the last term is suppressed, and we
recover the ordinary quantum cohomology relation~(\ref{eq:g24:qc:1-1}),
as expected.  (Ultimately this is due to a global $U(1)$ symmetry
present in two dimensions, but not three, resulting in a grading
respected by the quantum cohomology ring, but not the quantum
K theory ring.)  Proceeding in this fashion, it is straightforward to 
check that the $R \rightarrow 0$ limit of the quantum K theory relations
reproduces the quantum cohomology relations.

\section{Shifted Wilson line basis for Grassmannians}
\label{sect:gkn:shifted}

So far we have discussed bases consisting of 
Wilson lines (Schur polynomials in the $x_a$)
and Schubert classes (which must be computed on a case-by-case basis
for each geometry).  In this section, we will briefly discuss
another basis, shifted Wilson lines, which we will use for
computational efficiency later.

\subsection{Proposal}
\label{sect:gkn:shifted:prop}

Briefly, shifted Wilson lines (denoted $SW$) are Schur polynomials
in $z_a \equiv 1-x_a$, instead of the $x_a$.  They can be related
to a basis of Wilson lines with relatively straightforward algebra.
For example, for a $U(2)$ gauge theory, one has Wilson lines
\begin{eqnarray}
W_{\tiny\yng(1)} & = & x_1 + x_2,
\\
W_{\tiny\yng(2)} & = & x_1^2 + x_2^2 + x_1 x_2,
\\
W_{\tiny\yng(1,1)} & = & x_1 x_2.
\end{eqnarray}
In such a gauge theory, the shifted Wilson lines are defined by
\begin{eqnarray}
SW_{\tiny\yng(1)} & = & z_1 + z_2 \: = \: 2 - W_{\tiny\yng(1)},
\\
SW_{\tiny\yng(2)} & = & z_1^2 + z_2^2 + z_1 z_2 \: = \:
3 - 3 W_{\tiny\yng(1)} + W_{\tiny\yng(2)},
\\
SW_{\tiny\yng(1,1)} & = & z_1 z_2 \: = \:
1 - W_{\tiny\yng(1)} + W_{\tiny\yng(1,1)}.
\end{eqnarray}

We propose that
that the quantum K theory
relations for $G(k,n)$
can be described in a basis of shifted Wilson lines as follows.
(We plan to address the mathematical details in \cite{leom}.)
First, let $e_i(z)$ denote the elementary symmetric polynomials in $z$,
i.e.,
\begin{equation}
e_i(z) \: = \: SW_{1^i},
\end{equation}
where $SW_{1^i}$ denotes the shifted Wilson line associated to a Young tableau
consisting of one column with $i$ boxes, for example:
\begin{equation}
e_1(z) \: = \: \sum_a z_a, \: \: \:
e_2(z) \: = \: \sum_{a < b} z_a z_b,
\: \: \:
e_3(z) \: = \: \sum_{a < b < c} z_a z_b z_c,
\end{equation}
and so forth.  
(See e.g. \cite[appendix A]{Gu:2020oeb} for more information.)
Furthermore, in a basis of 
variables $v_1, \cdots, v_{n-k}$, we write
\begin{equation}  \label{eq:gkn:sw:vo}
e_i(v) \: = \: (-)^i {\cal O}_i,
\end{equation}
where ${\cal O}_i$ is the Schubert class associated to the Young tableau
with one row of $i$ boxes.
(We will see that the $v_{\ell}$ can be identified with roots distinct
from the $z_i$ of a characteristic polynomial arising from physics.)
Define $Z_i$, $V_i$ for $1 \leq i \leq n$ as follows:
\begin{equation}
Z_i \: = \: \left\{ \begin{array}{cl}
e_{i+1}(z) - e_{i+2}(z) + \cdots + (-)^{k-1-i} e_k(z)
& 1 \leq i \leq k-1, \\
0 & k \leq i \leq n,
\end{array} \right.
\end{equation}
\begin{equation}
V_i \: = \: \left\{ \begin{array}{cl}
0 & 1 \leq i \leq n-k, \\
(-)^{n-k} \left( \begin{array}{c} k-1 \\ n-i \end{array} \right) q &
n-k+1 \leq i \leq n.
\end{array} \right.
\end{equation}
Then,
\begin{equation}   \label{eq:gkn:sw:qk}
\sum_{a+b=i} (-)^b SW_{1^a} \star {\cal O}_b \: = \:
V_i \: + \:
\left\{ \begin{array}{cl}
SW_{1^{i+1}} - SW_{1^{i+2}} + \cdots +
(-)^{k-1-i} SW_{1^k} & 1 \leq i \leq k-1,
\\
0 & {\rm else}.
\end{array} \right.
\end{equation}
Rather than try to give a rigorous proof of this description
(see instead \cite{leom}), we will give a few examples,
plus a derivation from the physical chiral ring relations,
demonstrating that this basis is very natural for physics.

To be clear, the shifted Wilson line variables have appeared
previously in e.g. \cite{Jockers:2019lwe}; however,
we are not aware of previous work giving a presentation of the quantum
K theory ring in a basis of this form.

\subsection{Example:  projective space ${\mathbb P}^n$}

First, let us confirm that this gives the expected result for
the quantum K theory ring of a projective space
${\mathbb P}^n = G(1,n+1)$.
From the proposal above, we have $e_1(z) = SW_1 = z$,
$SW_{1^a} = 0$ for $a > 1$,
all $Z_i = 0$, and
\begin{equation}
V_i \: = \: \left\{ \begin{array}{cl}
0 & 1 \leq i \leq n,
\\
(-)^{n} q & i=n+1.
\end{array} \right.
\end{equation}
Furthermore, as noted
in section~\ref{sect:rev:pn}, for a projective space,
\begin{equation}
{\cal O}_1 \: = \: {\cal O}_{\tiny\yng(1)} \: = \:
{\cal O} - S \: = \: 1-x \: = \: z \: = \: SW_1,
\: \: \:
SW_0 \: = \: 1,
\end{equation}
and
\begin{equation}
{\cal O}_m \: = \: \left( {\cal O}_1 \right)^m \: = \: z^m.
\end{equation}
The quantum K theory ring is then predicted to have the 
relations~(\ref{eq:gkn:sw:qk})
\begin{equation}
SW_1 \star {\cal O} - SW_0 \star {\cal O}_1 \: = \:
0,
\end{equation}
since $SW_{1^2} = 0$, hence ${\cal O}_1 = SW_1$,
then for $2 \leq i \leq n$,
\begin{equation}
SW_1 \star {\cal O}_{i-1} - SW_0 \star {\cal O}_i
 \: = \:
0,
\end{equation}
hence ${\cal O}_i = SW_1 \star {\cal O}_{i-1}$,
and finally
\begin{equation}
(-)^n  SW_1 \star {\cal O}_n 
 \: = \: (-)^n q,
\end{equation}
or more simply, 
\begin{equation}
SW_1 \star {\cal O}_n \: = \: q,
\end{equation}
which algebraically is just $z^{n+1} = q$.
(We have used standard conventions in which the Schubert class
${\cal O}_{n+1} = 0$ on ${\mathbb P}^n$.)
This is precisely the expected quantum K theory relation for
${\mathbb P}^n$, so we see that the prediction~(\ref{eq:gkn:sw:qk})
does indeed work in this case.

\subsection{Example:  $G(2,4)$}

Next, let us turn to the case of $G(2,4)$.  Here, we will begin by rewriting
the physical ring relations in a way that will help link the details of the
mathematics proposal above, and then at the end we will give a detailed
computational verification of~(\ref{eq:gkn:sw:qk}) in this case.

In the case of $G(2,4)$, we can rewrite the physical ring 
relations~(\ref{eq:g24:eom1})
in terms of these variables as follows:
\begin{equation}
-q(1-z_1) \: = \: z_1^4 (1-z_2), \: \: \:
-q(1-z_2) \: = \: z_2^4 (1-z_1),
\end{equation}
or more simply,
\begin{equation}
z_a^4 - e_2(z) z_a^3 - q z_a + q \: = \: 0.
\end{equation}
Note that $z_{1,2}$ are two of the roots of the corresponding
polynomial
\begin{equation}
t^4 - e_2(z) t^3 - q t + q \: = \: 0.
\end{equation}
Let $w_{\ell}$ denote the four roots of this polynomial, then by comparing
coefficients with 
\begin{equation}
\prod_{\ell=1}^4 (t - w_{\ell})
 \: = \:
t^4 - e_1(w) t^3 + e_2(w) t^2 - e_3(w) t + e_4(w), 
\end{equation}
we have 
\begin{eqnarray}
e_1(w) & = & e_2(z), \\
e_2(w) & = & 0, \\
e_3(w) & = & q,  \label{eq:swg24:v3} \\
e_4(w) & = & q.  \label{eq:swg24:v4}
\end{eqnarray}
(This is often known more formally as Vieta's theorem.)
Without loss of generality we can identify $w_{1,2}$ with $z_{1,2}$,
then from the equations above, we have
\begin{eqnarray}
w_3 + w_4 & = & e_1(w) - e_1(z) \: = \: e_2(z) - e_1(z),
\\
w_3 w_4 & = & e_2(w) - e_2(z) - \left( z_1 w_3 + z_1 w_4 + z_2 w_3 + z_2 w_4
\right),
\\
& = & - e_2(z) - e_1(z) (w_3 + w_4) \: = \:
- e_2(z) - e_1(z) \left( e_2(z) - e_1(z) \right),
\\
& = & - e_2(z) - e_1(z) e_2(z) + e_1(z)^2. 
\end{eqnarray}

In passing, note that
if we identify $w_3$, $w_4$ with the $v_{\ell}$ in the statement of
the quantum K theory ring in shifted Wilson lines, then
\begin{eqnarray}
e_1(v) & = & w_3 + w_4 \: = \: e_2(z) - e_1(z) \: = \:
- \left( 1 - e_2(x) \right),
\\
& = & - {\cal O}_{\tiny\yng(1)},
\\
e_2(v) & = & w_3 w_4 \: = \: - e_2(z) - e_1(z) e_2(z) + e_1(z)^2,
\\
& = & 1 - 3 e_2(x) + e_1(x) e_2(x) 
\: = \: 
1 - 3 W_{\tiny\yng(1,1)} + W_{\tiny\yng(2,1)},
\\
& = & {\cal O}_{\tiny\yng(2)},
\end{eqnarray}
using the expressions for Schubert classes in
equations~(\ref{eq:g24:o1}), (\ref{eq:g24:o2}).
This confirms the dictionary~(\ref{eq:gkn:sw:vo}) for $G(2,4)$.

Plugging into equations~(\ref{eq:swg24:v3}), (\ref{eq:swg24:v4}), 
we have
\begin{equation} \label{eq:sw:g24:1}
e_1(z)^3 - 2 e_1(z) e_2(z) - e_1(z)^2 e_2(z) + e_2(z)^2 \: = \: q,
\end{equation}
\begin{equation}  \label{eq:sw:g24:2}
e_1(z)^2 e_2(z) - e_2(z)^2 - e_1(z) e_2(z)^2 \: = \: q.
\end{equation}
These are the implications of the physical ring relations for
$G(2,4)$, in the shifted variables.
It is straightforward to check that 
equation~(\ref{eq:sw:g24:1}) matches 
equation~(\ref{eq:g24-eom-res2}) from our previous analysis of the GLSM for
$G(2,4)$.
For later comparisons, note that the difference between
these two equations is
\begin{equation}  \label{eq:sw:g24:diff}
\left( e_1(z) - e_2(z) \right)
\left( e_1(z)^2 - 2 e_2(z) - e_1(z) e_2(z) \right) \: = \: 0.
\end{equation}
In $x$ variables, this becomes
\begin{equation}
W_{\tiny\yng(1,1)} \cdot W_{\tiny\yng(2,1)} \: = \:
W_{\tiny\yng(1)} - 4 W_{\tiny\yng(1,1)} + 4 W_{\tiny\yng(2,2)},
\end{equation}
which matches the product~(\ref{eq:g24:21-11}) derived earlier.

Now, to close our discussion of $G(2,4)$, let us give a detailed
verification of the proposal~(\ref{eq:gkn:sw:qk}) in this example.
Briefly, that proposal predicts the following relations:
\begin{enumerate}
\item \begin{equation}
SW_1 \star {\cal O} - SW_0 \star {\cal O}_1 \: = \: SW_{1^2},
\end{equation}
or more simply,
\begin{equation}
{\cal O}_{\tiny\yng(1)} \: = \: e_1(z) - e_2(z).
\end{equation}
\item \begin{equation}
SW_{1^2} \star {\cal O} - SW_1 \star {\cal O}_1 + SW_0 \star {\cal O}_2
\: = \: 0,
\end{equation}
or more simply,
\begin{equation}
{\cal O}_{\tiny\yng(2)} \: = \: {\cal O}_2 \: = \:
 e_1(z) {\cal O}_{\tiny\yng(1)} - e_2(z).
\end{equation}
\item \begin{equation}
- SW_{1^2} \star {\cal O}_1 + SW_1 \star {\cal O}_2 \: = \: q,
\end{equation}
or more simply,
\begin{equation}
- e_2(z) {\cal O}_{\tiny\yng(1)} \: + \: e_1(z) {\cal O}_{\tiny\yng(2)}
\: = \: q,
\end{equation}
\item \begin{equation}
SW_{1^2} \star {\cal O}_2 \: = \: q.
\end{equation}
or more simply,
\begin{equation}
e_2(z) {\cal O}_{\tiny\yng(2)} \: = \: q.
\end{equation}
\end{enumerate}
The first two equations are equivalent to the definitions of
${\cal O}_{\tiny\yng(1)}$ and ${\cal O}_{\tiny\yng(2)}$.
The second two equations, 
match the Vieta theorem implications~(\ref{eq:sw:g24:1}), (\ref{eq:sw:g24:2}).
Thus, we see that the proposal~(\ref{eq:gkn:sw:qk}) correctly reproduces
the physics predictions for quantum K theory relations in shifted Wilson
line variables for $G(2,4)$.

\subsection{Derivation for general cases}

Now that we have seen how the proposal works in a few examples,
we will give a general argument for why the 
proposal~(\ref{eq:gkn:sw:qk}) arises from physics.

The equations of motion are given in equation~(\ref{eq:gkn-eom}),
\begin{equation} \label{eq:gkn-eom2}
(-)^{k-1} q x_a^{k-1} \: = \: (1-x_a)^n
\left( \prod_{b\neq a} x_b \right) 
\end{equation}
(after a cancellation following from the fact that the $x_a \neq 0$).
In shifted Wilson line variables $z_a = 1-x_a$, this is
\begin{eqnarray} 
(-)^{k-1} q (1 - z_a)^{k-1} & = &
z_a^n \left( \prod_{b \neq a} (1-z_b) \right),
\\
& = & z_a^n \left[ 1 \: - \: \left( \sum_{b \neq a} z_b \right)
\: + \:
\left( \sum_{b < c, b, c \neq a} z_b z_c \right) \: - \: \cdots \right].
\end{eqnarray}

We can rewrite the left-hand side in terms of Weyl-invariant combinations
by multiplying in factors of $z_a$, and successively adding/subtracting
terms.
For example, for $G(3,6)$, for $a=1$, the left-hand side is
\begin{eqnarray}
z_1^6 \left[ 1 - (z_2 + z_3) + (z_2 z_3) \right]
& = &
z_1^6 - z_1^5( z_1 z_2 + z_1 z_3) + z_1^5 (z_1 z_2 z_3),
\\
& = & z_1^6 - z_1^5 e_2(z) + z_1^5 z_2 z_3 + z_1^5 e_3(z),
\\
& = &
z_1^6 - z_1^5 \left( e_2(z) - e_3(z) \right) + z_1^4 e_3(z).
\end{eqnarray}
Proceeding in this fashion, it is straightforward to demonstrate that
\begin{eqnarray}
\lefteqn{
z_a^n \left( \prod_{b \neq a} (1-z_b) \right)
} \nonumber \\
& \hspace*{0.25in} = & z_a^n \: - \: z_a^{n-1} \left( e_2(z) - e_3(z) + e_4(z) - \cdots
+ (-)^{k-2} e_k(z) \right)
\nonumber \\
& & \hspace*{0.3in}
\: + \:
z_a^{n-2} \left( e_3(z) - e_4(z) + e_5(z) - \cdots + (-)^{k-3} e_k(z)
\right)
\nonumber \\
& & \hspace*{0.3in}
\: - \:
z_a^{n-3} \left( e_4(z) - e_5(z) + e_6(z) - \cdots + (-)^{k-4} e_k(z) \right)
\nonumber \\
& &  \hspace*{0.7in}
\: + \: \cdots \: + \:
(-)^{n-k+1} z_a^{n-k+1} e_k(z).   \label{eq:gkn:sw:piece1}
\end{eqnarray}

Thus, the physics equations of motion become
\begin{eqnarray}
(-)^{k-1} q (1 - z_a)^{k-1} 
& = &
z_a^n \: - \: z_a^{n-1} \left( e_2(z) - e_3(z) + e_4(z) - \cdots
+ (-)^{k-2} e_k(z) \right)
\nonumber \\
& & \hspace*{0.3in}
\: + \:
z_a^{n-2} \left( e_3(z) - e_4(z) + e_5(z) - \cdots + (-)^{k-3} e_k(z)
\right)
\nonumber \\
& & \hspace*{0.3in}
\: - \:
z_a^{n-3} \left( e_4(z) - e_5(z) + e_6(z) - \cdots + (-)^{k-4} e_k(z) \right)
\nonumber \\
& &  \hspace*{0.7in}
\: + \: \cdots \: + \:
(-)^{n-k+1} z_a^{n-k+1} e_k(z).
\end{eqnarray}
Since this holds for all values of $a$, we can think of the $a$ as 
roots of the following polynomial, which we will refer to as a
characteristic polynomial:
\begin{eqnarray}
(-)^{k-1} q (1 - t)^{k-1} 
& = &
t^n \: - \: t^{n-1} \left( e_2(z) - e_3(z) + e_4(z) - \cdots
+ (-)^{k-2} e_k(z) \right)
\nonumber \\
& & \hspace*{0.3in}
\: + \:
t^{n-2} \left( e_3(z) - e_4(z) + e_5(z) - \cdots + (-)^{k-3} e_k(z)
\right)
\nonumber \\
& & \hspace*{0.3in}
\: - \:
t^{n-3} \left( e_4(z) - e_5(z) + e_6(z) - \cdots + (-)^{k-4} e_k(z) \right)
\nonumber \\
& &  \hspace*{0.7in}
\: + \: \cdots \: + \:
(-)^{n-k+1} t^{n-k+1} e_k(z).
\end{eqnarray}

In passing, although we have used the term `characteristic polynomial,'
it is important to note that it is not unique.  For example,
the polynomial $z_1 (z_1 + z_2)$  can be interpreted as either
$z_1^2 + e_2$ or $z_1 e_1$,
which lead to $t^2 + e_2$ or $t e_1$, respectively.  Regardless of
choices, the desired $z_a$ will still emerge as some of the roots.
That said, the interpretation of the $e_i(v)$ (for $v$ denoting the
remaining roots) may change.

The characteristic polynomial above has degree $n$, 
but there are only $k$ $z_a$'s.
We will denote the other $n-k$ roots by $v_{\ell}$, for $1 \leq \ell \leq n-k$.
Let $\{w_i\}$ 
denote the collection of all roots $\{ z_a, v_{\ell} \}$.
Since the coefficient of $t^n$ in the characteristic polynomial is $1$,
we can write it as
\begin{equation}
\prod_i ( t - w_i) \: = \: t^n \: - \:
e_1(w) t^{n-1} \: + \: e_2(w) t^{n-2} \: - \:
e_3(w) t^{n-3} \: + \: \cdots \: + \: (-)^n e_n(w).
\end{equation}
Clearly, the coefficient of $t^{n-i}$ is $(-)^i e_i(w)$.
Comparing to the characteristic polynomial above,
the coefficient of $t^{n-i}$ in the right-hand-side is $(-)^i Z_i$,
in the notation of the proposal~(\ref{eq:gkn:sw:qk}),
and similarly the coefficient of $t^{n-i}$ in 
\begin{equation}
- (-)^{k-1} q (1-t)^{k-1}
\end{equation}
is $(-)^i V_i$.  Putting this together, from comparing
the coefficients of $t^{n-i}$,
and cancelling out the common $(-)^i$ factor, we have
\begin{equation}
e_i(w) \: = \: Z_i + V_i.
\end{equation}

Now, 
it is straightforward to see that
\begin{eqnarray}
e_1(w) & = & e_1(z) + e_1(v),
\\
e_2(w) & = & e_2(z) + e_1(z) e_1(v) + e_2(v),
\\
e_3(w) & = & e_3(z) + e_2(z) e_1(v) + e_1(z) e_2(v) + e_3(v),
\end{eqnarray}
and so forth.
Furthermore,
\begin{equation}
e_i(w) \: = \: SW_{1^i},
\end{equation}
and, as proposed in section~\ref{sect:gkn:shifted:prop},
\begin{equation}
e_j(v) \: = \: (-)^j {\cal O}_j,
\end{equation}
hence
\begin{equation}
e_i(w) \: = \: \sum_{a+b=i} (-)^b SW_{1^a} \star {\cal O}_b.
\end{equation}
Assembling these pieces, we have
\begin{equation}
 \sum_{a+b=i} (-)^b SW_{1^a} \star {\cal O}_b
\: = \: V_i + Z_i,
\end{equation}
which is the proposal~(\ref{eq:gkn:sw:qk}) for the quantum K theory ring
of $G(k,n)$.

\subsection{Useful identities}
\label{sect:useful-identities}

Finally, let us close this subsection
with some identities that will be helpful later.
Let $e_i(x)$ denote the $i$th elementary symmetric polynomial in the
$x_a$, as before.  
Then, for gauge group $U(k)$ (so that there are $k$ $x_a$'s),
\begin{eqnarray}
e_1(x) & = & \sum_a x_a \: = \: \sum_a (1 - z_a) \: = \:
k - e_1(z),
\\
e_2(x) & = & \sum_{a < b} x_a x_b \: = \:
\left( \begin{array}{c} k \\ 2 \end{array} \right)
\: - \: \left( \begin{array}{c} k-1 \\ 1 \end{array} \right) e_1(z)
\: + \: e_2(z),
\\
e_3(x) & = & \sum_{a < b < c} x_a x_b x_c ,
\\
 & = &
\left( \begin{array}{c} k \\ 3 \end{array} \right)
\: - \:
\left( \begin{array}{c} k-1 \\ 2 \end{array} \right) e_1(z)
\: + \:
\left( \begin{array}{c} k-2 \\ 1 \end{array} \right) e_2(z)
\: - \: e_3(z),
\end{eqnarray}
and more generally, for $n \leq k$,
\begin{equation}
e_n(x) \: = \:
\left( \begin{array}{c} k \\ n \end{array} \right)
\: - \:
\left( \begin{array}{c} k-1 \\ n-1 \end{array} \right) e_1(z)
\: + \:
\left( \begin{array}{c} k-2 \\ n-2 \end{array} \right) e_2(z)
\: + \: \cdots \: + \: (-)^n e_n(z).
\end{equation}

\section{$\lambda_y$ class relations for ordinary Grassmannians}
\label{sect:lambda:gkn}

In this section we will give another description of the quantum K theory
ring of Grassmannians, 
in a basis determined by the exterior powers of the universal subbundle and
universal quotient bundle.  An analogous description will play an
important later in relating the quantum K theory of Lagrangian
Grassmannians to physics computations.

To this end, 
for any vector bundle ${\cal E} \rightarrow X$ of rank $r$,
define
\begin{equation}
\lambda_y({\cal E}) \: = \: 1 \oplus y {\cal E} \oplus
y^2 \wedge^2 {\cal E} \oplus \cdots \oplus y^r \wedge^r {\cal E}
\end{equation}
as an element of K theory.
One of the basic properties of this construction is that
\begin{equation}
\lambda_y( {\cal E} \oplus {\cal F}) \: = \: \left( \lambda_y {\cal E} \right)
\otimes \left( \lambda_y {\cal F} \right).
\end{equation}
(See e.g. \cite[appendix A]{Ando:2009av} for other useful identities satisfied
by this and the related symmetrization map.)
In particular, applying the splitting principle, if (formally) we write
${\cal E}$ as a sum of line bundles
\begin{equation}
{\cal E} \: = \: {\cal L}_1 \oplus \cdots \oplus {\cal L}_r,
\end{equation}
then
\begin{equation}
\lambda_y {\cal E} \: = \: \bigotimes_a \left( 1 + y {\cal L}_a \right).
\end{equation}

Now, let us apply this to give a description of the quantum K theory
ring of a Grassmannian.
Begin with the canonical short exact sequence on
a Grassmannian $G(k,n)$ relating the (rank $k$) universal subbundle
$S$ to the (rank $n-k$) universal quotient bundle $Q$:
\begin{equation}
0 \: \longrightarrow \: S \: \longrightarrow \: 
{\cal O}^n \: \longrightarrow \: Q \: \longrightarrow \: 0.
\end{equation}
Classically, this implies that
\begin{equation}
\lambda_y(S) \lambda_y(Q) \: = \: (1+y)^n,
\end{equation}
where we identify $1 = {\cal O}$.

We propose that
the quantum K theory relations 
for $G(k,n)$ are given by
\begin{eqnarray}  \label{eq:lambda:gkn}
\lambda_y(S) \star \lambda_y(Q) & = & (1+y)^n \: - \:
q \sum_{i=0}^{k-1} y^{n-i} \wedge^i S^*,
\\
& = &
(1+y)^n \: - \:
q (\det Q) \otimes y^{n-k} \left( \lambda_y(S) - 1 \right),
\\
\lambda_y(Q^*) \star \lambda_y(S^*) & = &
(1+y)^n \: - \:
q \sum_{i=1}^{n-k-1} y^{n-i} \wedge^i Q,  \label{eq:lambda:gkn2}
\\
\wedge^k S & = & \wedge^{n-k} Q^*,  \label{eq:lambda:gkn3}
\\
\wedge^k S^* & = & \wedge^{n-k} Q,   \label{eq:lambda:gkn4}
\end{eqnarray}
where $\star$ denotes the product in quantum K theory, and
$\otimes$ is the classical tensor product.
(A mathematical proof of this relation will appear in \cite{leom}.)
The reader should note that the relations are exchanged by
the duality $G(k,n) = G(n-k,n)$, which exchanges $S^*$ and $Q$.
The quantum K theory ring of $G(k,n)$ is then
\begin{equation}
{\mathbb Z}[q][X_1, \overline{X}_1, X_2, \cdots, \overline{X}_k,
Y_1, \overline{Y}_1, \cdots, \overline{Y}_{n-k} ] / J,
\end{equation}
where $J$ is the ideal generated by
\begin{eqnarray}
\lefteqn{
\left( 1 + y X_1 + y^2 X_2 + \cdots + y^k X_k \right)
\left( 1 + y Y_1 + y^2 Y_2 + \cdots + y^{n-k} Y_{n-k} \right)
} \nonumber \\
& \hspace*{2in} = &
(1 + y)^n - q \sum_{i=0}^{k-1} y^{n-i} \overline{X}_i,
\end{eqnarray}
\begin{eqnarray}
\lefteqn{
\left( 1 + y \overline{Y}_1 + y^2 \overline{Y}_2 + \cdots +
 y^{n-k} \overline{Y}_{n-k} \right)
\left( 1 + y \overline{ X}_1 + y^2 \overline{X}_2 + \cdots +
 y^k \overline{X}_k \right)
} \nonumber \\
& \hspace*{2in} = &
(1 + y)^n - q \sum_{i=0}^{n-k-1} y^{n-i} Y_i,
\end{eqnarray}
\begin{eqnarray}
X_k & = & \overline{Y}_{n-k},
\\
\overline{X}_k & = & Y_{n-k},
\end{eqnarray}
corresponding to the  $\lambda_y$
relations~(\ref{eq:lambda:gkn})-(\ref{eq:lambda:gkn4}) above.

This description of the quantum K theory ring is motivated by
an analogous description of the ordinary quantum cohomology of
a Grassmannian $G(k,n)$, given in e.g. \cite[section 3.2]{Witten:1993xi} as
\begin{equation}
c(S^*) c(Q^*) \: = \: 1 + (-)^{n-k} q,
\end{equation}
where $q$ is taken to have cohomological degree $n$.  In particular,
the $q$ correction itself is encoded in 
\begin{equation}  \label{eq:gkn:qc:2}
c_k(S^*) c_{n-k}(Q^*) \: = \: (-)^{n-k} q.
\end{equation}

In fact, we can derive the quantum cohomology relation~(\ref{eq:gkn:qc:2})
as a two-dimensional limit of the quantum K theory 
relation~(\ref{eq:lambda:gkn}),
as follows.  To do this, set $y=-1$, for which
\begin{equation}
\lim_{R \rightarrow 0} \lambda_{-1}(S) \: \sim \: (-)^k R^k c_k(S),
\: \: \:
\lim_{R \rightarrow 0} \lambda_{-1}(Q) \: \sim \: (-)^{n-k} R^{n-k} c_{n-k}(Q),
\end{equation}
\begin{equation}
\lim_{R \rightarrow 0} \det Q \: = \:
\lim_{R \rightarrow 0} \prod_{\ell} \tilde{x}_{\ell} \: = \: 1,
\end{equation}
where without loss of generality we have suppressed factors of $2 \pi i$,
and use equation~(\ref{eq:3d2d:funds}) to relate $q_{3d}$ to $R^n q_{2d}$,
so that the quantum K theory relation~(\ref{eq:lambda:gkn}) reduces
in the limit $R \rightarrow 0$ to
\begin{equation}
(-)^n R^n c_k(S) c_{n-k}(Q) \: = \: 0 - R^n q_{2d} (-)^{n-k} \left(
-1 + (-)^k R^k c_k(S) \right).
\end{equation}
Dividing out common $R^n$ factors and suppressing subleading terms which 
vanish in the limit, we
have
\begin{equation}
c_k(S) c_{n-k}(Q) \: = \:  (-)^{-k} q_{2d},
\end{equation}
or equivalently
\begin{equation}
c_k(S^*) c_{n-k}(Q^*) \: = \:  (-)^{n-k} q_{2d},
\end{equation}
which matches~(\ref{eq:gkn:qc:2}).

Now,
let us examine this quantum K theory relation in detail for a projective space
${\mathbb P}^n = G(1,n+1)$.
The relation~(\ref{eq:lambda:gkn}) reduces to
\begin{equation}
\lambda_y(S) \star \lambda_y(Q) \: = \: (1+y)^{n+1} - q y^{n+1},
\end{equation}
since classically $S \cong (\det Q)^{-1}$.  We can expand this out by
coefficients of $y$ as follows:
\begin{eqnarray}
S + Q & = & n+1,
\\
S \star Q + \wedge^2 Q & = & \left( \begin{array}{c} n+1 \\ 2 \end{array}
\right),
\\
S \star \wedge^2 Q + \wedge^3 Q & = & \left( \begin{array}{c} n+1 \\ 3 
\end{array} \right),
\end{eqnarray}
and so forth, culminating in
\begin{eqnarray}
S \star \wedge^{n-1} Q + \wedge^n Q & = & \left( \begin{array}{c} n+1 \\ n
\end{array}  \right),
\\
S \star \wedge^n Q & = & 1-q,
\end{eqnarray}
where we have identified $1$ with ${\cal O}$.
Solving these equations iteratively, we find
\begin{eqnarray}
Q & = & n+1 - S,
\\
\wedge^2 Q & = & \left( \begin{array}{c} n+1 \\ 2 \end{array} \right)
- (n+1) S + S \star S,
\\
\wedge^3 Q & = & \left( \begin{array}{c} n+1 \\ 3 \end{array} \right)
- \left( \begin{array}{c} n+1 \\ 2 \end{array} \right) S +
(n+1) S \star S - S \star S \star S,
\end{eqnarray}
and so forth, with the last equation, arising from the coefficient of
$y^{n+1}$, implying
\begin{eqnarray}
q & = & 1 - S \star \wedge^n Q,
\\
& = & 1 - S \left( \begin{array}{c} n+1 \\ 1 \end{array} \right)
+ S^2 \left( \begin{array}{c} n+1 \\ 2 \end{array} \right)
+ \cdots + (-)^{n+1} S^{n+1},
\\
& = & (1-S)^{n+1},
\end{eqnarray}
where we have used $S^n$ to denote the quantum ($\star$) product of
$n$ copies of $S$.
This coincides with the quantum K theory relation~(\ref{eq:pn:qk}) for
${\mathbb P}^n$, where we identify $x$ with $S$.

Before going on, let us recast these conventions in a slightly different
notation which will make more explicit the connection to gauged linear
sigma models.  If we apply the splitting principle to formally write
$S = \oplus_a x_a$, then
we can write
\begin{eqnarray}
\lambda_y(S) & = & 1 + y S + y^2 \wedge^2 S + y^3 \wedge^3 S + \cdots,
\\
& = & 1 + y e_1(x) + y^2 e_2(x) + y^3 e_3(x) + \cdots,
\end{eqnarray}
where the $e_a(x)$ are the elementary symmetric polynomials
in the
$x_a$, for example:
\begin{equation}
e_1(x) \: = \: \sum_a x_a, \: \: \:
e_2(x) \: = \: \sum_{a < b} x_a x_b,
\: \: \:
e_3(x) \: = \: \sum_{a < b < c} x_a x_b x_c,
\end{equation}
and so forth, as before.  
These $x_a$'s associated to $S$ can be identified with 
the quantities appearing in the GLSM twisted one-loop effective
superpotential, the exponentials of the $\sigma_a$.
Similarly, applying the splitting principle we can formally write
$Q = \oplus_{\ell} \tilde{x}_{\ell}$, so that similarly
\begin{equation}
\lambda_y(Q) \: = \: 1 + y e_1(\tilde{x}) + y^2 e_2(\tilde{x}) +
y^3 e_3(\tilde{x}) + \cdots,
\end{equation}
but in general the $\tilde{x}_{\ell}$ are not directly connected to
GLSM variables.  (One prominent exception we will discuss later will
be $LG(n,2n)$, for which $Q = S^*$, so we can choose $\tilde{x}_{\ell}$
to be $x_a^{-1}$, as we will discuss.)

Now, let us turn to $G(2,4)$.
Here, the relation~(\ref{eq:lambda:gkn}) reduces to
\begin{eqnarray}
\lambda_y(S) \star \lambda_y(Q) & = & (1+y)^4 - q (\det Q) \otimes
\left( y^{3} S + y^4 \wedge^2 S \right),
\\
& = & (1+y)^4 - q \left( y^{3} S \otimes (\det Q) + y^4 \right),
\end{eqnarray}
using the fact that classically $\wedge^2 S \cong (\det Q)^{-1}$.
Expanding in powers of $y$, we get the relations
\begin{eqnarray}
S + Q & = & 4, 
\\
\wedge^2 S + S \star Q + \wedge^2 Q & = & 6,
\\
\wedge^2 S \star Q + S \star \wedge^2 Q & = & 4 - q \, S \otimes (\det Q)
\: = \: 4 - q S^*,
\\
\wedge^2 S \star \wedge^2 Q & = & 1 - q,
\end{eqnarray}
or equivalently, in terms of elementary symmetric polynomials,
\begin{eqnarray}
e_1(x) + e_1(\tilde{x}) & = & 4,
\\
e_2(x) + e_1(x) e_1(\tilde{x}) + e_2(\tilde{x}) & = & 6,
\\
e_2(x) e_1(\tilde{x}) + e_1(x) e_2(\tilde{x}) & = &
4 - q \, e_1(S^*),
\\
e_2(x) e_2(\tilde{x}) & = & 1 - q,
\end{eqnarray}
where we have used the fact that classically 
$S \otimes \det Q = S^*$ for the
case of $G(2,4)$, and that 
\begin{equation}
e_1(S^*) \: = \: 
2 \: + \: {\cal O}_{\tiny\yng(1)} + {\cal O}_{\tiny\yng(2)} 
\: = \:
4 - 4 e_2(x) + e_1(x) e_2(x).
\end{equation}
(In particular, $S^* \neq Q$, as they have different $c_2$'s.)

In passing, the reader may find it helpful to note that for
any vector bundle $E$, which by the splitting theorem
$E = \oplus_a x_a$ formally,
\begin{equation}
e_m(x) \: = \: e_m(E) \: = \: e_1(\wedge^m E).
\end{equation}

Now, let us simplify the expressions above.
We can use the first two equations to eliminate $e_1(\tilde{x})$ and
$e_2(\tilde{x})$:
\begin{eqnarray}
e_1(\tilde{x}) & = & 4 - e_1(x),
\\
e_2(\tilde{x}) & = & 6 - e_2(x) - 4 e_1(x) + e_1(x)^2,
\end{eqnarray}
or equivalently,
\begin{eqnarray}
Q & = & 4 {\cal O} - S,
\\
\wedge^2 Q & = & 6 {\cal O} - \wedge^2 S - 4 S + S^2.
\end{eqnarray}
Plugging into the second two equations, we have
\begin{eqnarray}
\lefteqn{
6 e_1(x) + 4 e_2(x) - 2 e_1(x) e_2(x) - 4 e_1(x)^2 + e_1(x)^3
} \nonumber \\
 & \hspace*{1in} = &
4 - 4 q + 4 q e_2(x) - q e_1(x) e_2(x),    \label{eq:g24:lambda:1}
\end{eqnarray}
\begin{eqnarray}
6 e_2(x) - e_2(x)^2 - 4 e_1(x) e_2(x) + e_1(x)^2 e_2(x) & = & 1 - q.
\label{eq:g24:lambda:2}
\end{eqnarray}

So far, we have merely made explicit the implications of the $\lambda_y$
class relations.  We still need to compare to physics predictions.
To do so, we will first move to the shifted Wilson line basis
of section~(\ref{sect:gkn:shifted}.  Using the dictionary
\begin{equation}
e_1(x) \: = \: 2 - e_1(z), \: \: \:
e_2(x) \: = \: 1 - e_1(z) + e_2(z),
\end{equation}
the $\lambda_y$ class relations~(\ref{eq:g24:lambda:1}), 
(\ref{eq:g24:lambda:2}) become
\begin{eqnarray}
\lefteqn{
2 e_1(z) e_2(z) - e_1(z)^3
} \nonumber \\
 & \hspace*{0.5in} = & - q \left( 2 + e_1(z) + e_1(z)^2 -
2 e_2(z) - e_1(z) e_2(z) \right),
\label{eq:lambda:g24:1}
\end{eqnarray}
\begin{eqnarray}
- e_2(z)^2 - e_1(z)^3 + 2 e_1(z) e_2(z) + e_1(z)^2 e_2(z)
& = & -q.
\label{eq:lambda:g24:2}
\end{eqnarray}

Now, let us compare these equations to physics.
The reader should first note that the second equation above,
(\ref{eq:lambda:g24:2}), is the same as the first of the
shifted Wilson line expressions~(\ref{eq:sw:g24:1}).
Next, if we eliminate $q$, we can write the first equation as
\begin{eqnarray}
\lefteqn{
\left[ - e_2(z)^2 - e_1(z)^3 + 2 e_1(z) e_2(z) + e_1^2(z) e_2(z) \right]
\left[ 2 + e_1(z) + e_1(z)^2 - 2 e_2(z) - e_1(z) e_2(z) \right]
} \nonumber \\
&\hspace*{4in} = & 2 e_1(z) e_2(z) - e_1(z)^3  \nonumber
\end{eqnarray}
or more simply
\begin{equation}
\left( e_1(z) - e_2(z) \right)
\left[ e_1(z)^2 - 2 e_2(z) - e_1(z) e_2(z) \right]
\left[ -1 - e_1(z)  e_1(z)^2 + e_2(z) \right] \: = \: 0.
\end{equation}
The first two factors were shown to vanish from the physics 
relation~(\ref{eq:sw:g24:diff}), hence we see that the physics relations
imply the $\lambda_y$ class relations for $G(2,4)$.

\section{Hypersurfaces in projective space}
\label{sect:levels}

So far, we have reviewed how physical Coulomb branch considerations yield
mathematical quantum K theory rings in three-dimensional theories
without a superpotential.   
In this section we will outline some of the complications that
ensue when one adds a superpotential, in the case of hypersurfaces
in projective spaces, as a warm-up before considering the quantum K theory
ring of symplectic Grassmannians.  (See also 
\cite[section 2.1]{Jockers:2018sfl} for a related
discussion in terms of I-functions.)

\subsection{Generalities}
\label{sect:hyp:gen}

Consider $U(1)$ gauge theories describing hypersurfaces of
degree $d$ in a projective space ${\mathbb P}^n$.  If there were
no superpotential, if we were describing the total space of
the line bundle ${\cal O}(-d) \rightarrow {\mathbb P}^n$,
then we would use the ansatz described earlier and take
\begin{equation}
k \: = \: - \frac{1}{2} \left( n + 1 + d^2 \right).
\end{equation}
The twisted one-loop effective superpotential is then
\begin{eqnarray}
W & = & \frac{k}{2} \left( \ln x \right)^2 \: + \: (\ln q) (\ln x)
\: + \:
(n+1) {\rm Li}_2( x ) \: + \: {\rm Li}_2\left( x^{-d} \right)
\nonumber \\
& & 
\: + \: \frac{n+1}{4} (\ln x)^2 \: + \: \frac{1}{4} (- d \ln x)^2,
\\
& = &
  (\ln q) (\ln x) \: + \:
(n+1) {\rm Li}_2( x ) \: + \: {\rm Li}_2\left( x^{-d} \right).
\end{eqnarray}
The physical chiral ring relation for this noncompact model is
\begin{equation}
(1 - x)^{n+1}
\: = \:
q \left( 1 - x^{-d}\right)^{d}.
\end{equation}

When we add a superpotential, so as to describe the compact
hypersurface, we take instead
\begin{equation}
k \: = \:  - \frac{1}{2} \left( n + 1 + d^2 \right) \: + \: d^2
\: = \:  - \frac{1}{2} \left( n + 1 - d^2 \right).
\end{equation}
The twisted one-loop effective superpotential is then
\begin{eqnarray}
W & = &
\frac{d^2}{2} ( \ln x )^2 \: + \:
 (\ln q) (\ln x) \: + \:
(n+1) {\rm Li}_2( x ) \: + \: {\rm Li}_2\left( x^{-d} \right),
\end{eqnarray}
from which one derives the
the chiral ring relation
\begin{equation}  \label{eq:chiralring:pn-d}
(1 - x)^{n+1} \: = \:
(-)^d q \left(1 - x^d \right)^{d}.
\end{equation}
(See also \cite[equ'n (2.24)]{Jockers:2018sfl}.
The corresponding quantum cohomology ring arising in the $R \rightarrow 0$
limit also agrees with that appearing in
\cite[equ'n (1.1)]{Collino:1996my}.)
In effect, the $x$ appearing in the $p$ field factor has been
replaced by $1/x$.

We observe that for
$U(1)$ theories, a more general formula that encompasses such cases
is
\begin{equation}
k \: = \: - (1/2) \sum_i Q_i |Q_i| .
\end{equation} 

Next, let us briefly consider the possibility of
topological vacua.  If these exist, they are
located at the vanishing locus of a function.
In the notation of \cite{Intriligator:2013lca}
(see also \cite{Bullimore:2019qnt}),
for massless chirals, that function is
\begin{equation}
F(\sigma) \: = \: \zeta + k \sigma + \frac{1}{2} \left(  \sum_i Q_i |Q_i|
\right) | \sigma |,
\end{equation}
where $\zeta$ is the Fayet-Iliopoulos parameter.
For the $k$ above, this simplifies to
\begin{equation}
F(\sigma) \: = \: \zeta + k \left(  \sigma - | \sigma | \right).
\end{equation}
Note that for $\zeta > 0$,
this theory has no topological vacua, since $F$ never vanishes.

Before going on, let us elaborate on why we chose the level above
for the compact hypersurface.  One way to motivate this is through
comparing the K-theoretic I functions, as in
\cite[parts 4, 5]{g11}, \cite[section 2]{Jockers:2019wjh}.
The I function for
a degree $\ell$ hypersurface in ${\mathbb P}^n$ has the form
\begin{equation}
\sum_{d \geq 0} \frac{
\prod_{k=1}^{\ell d} (1 - x^{\ell} Q^k )
}{
\prod_{k=1}^d (1 - x Q^k )^{n+1}
} q^d,
\end{equation}
where $Q = \exp(- \beta \hbar)$, $\beta$ determined by the radius
of the $S^1$, and $x$ an exponential of $\sigma$. 
This corresponds to a chiral ring relation
\begin{equation}
(1 - x)^{n+1} \: = \: q (1 - x^{\ell})^{\ell}.
\end{equation} 
The I function for the corresponding $V_+$ model, the total space
of the line bundle ${\cal O}(-\ell) \rightarrow {\mathbb P}^n$, is given by
\begin{equation}
\sum_{d \geq 0} \frac{
\prod_{k=0}^{\ell d - \ell} (1 - x^{-\ell} Q^k )
}{
\prod_{k=1}^d (1 - x Q^k )^{n+1}
} q^d,
\end{equation}
and the corresponding chiral ring relation is
\begin{equation}
(1 - x)^{n+1} \: = \: q (1 - x^{-\ell})^{\ell}.
\end{equation}
The difference between them is the power of $x$ appearing in the numerator,
which in a GLSM corresponds to the R-charge $2$ $p$ field.
This suggests that to describe a compact hypersurface instead of the
noncompact total space of a vector bundle, we need to add a contribution to
the level that algebraically inverts the $x$'s corresponding to the $p$
field, which is mechanically the difference between the Chern-Simons
levels for a noncompact $V_+$ model and a corresponding compact hypersurface in 
a projective space.  We will use the same method to arrive at a proposal
for Chern-Simons levels suitable for symplectic Grassmannians later.

\subsection{Degree one hyperplanes}

Mathematically, a linear hypersurface in ${\mathbb P}^n$ is
just ${\mathbb P}^{n-1}$.  We can see this in physics as follows.
Consider a supersymmetric $U(1)$ gauge theory with $n+1$ chiral
multiplets $x_i$ of charge $+1$, and one chiral multiplet $p$ of
charge $-1$, with superpotential
\begin{equation}
W \: = \: p x_{n+1}.
\end{equation}
The superpotential acts as a mass term, removing both $p$ and $x_{n+1}$,
leaving at lower energies a $U(1)$ gauge theory with $n$ chiral
multiplets of charge $+1$ and no superpotential, describing
${\mathbb P}^{n-1}$.

The ordinary quantum cohomology rings behave similarly.
The ring relation has the form
\begin{equation}
\sigma^{n+1} \sigma^{-1} \: = \: q,
\end{equation}
from which one immediately reads off $\sigma^n = q$, as expected.

Now, let us consider the quantum K theory relations.
From the general formula~(\ref{eq:chiralring:pn-d}) for
${\mathbb P}^n[d]$ for the case $d=1$, we get
\begin{equation}
(1 - x)^{n+1} \: = \: - q (1 - x)
\end{equation}
or more simply,
\begin{equation}
(1 - x)^n \: = \: -q,
\end{equation}
which is the quantum K theory ring relation for ${\mathbb P}^{n-1}$.

\subsection{Degree two hypersurfaces}

Let us first recover the ordinary quantum cohomology
ring of ${\mathbb P}^n[2]$ from a GLSM, from Coulomb branch considerations.
The GLSM has gauge group $U(1)$, with $n+1$ chiral superfields $\phi_i$
of charge
$1$, one chiral superfield $p$ of charge $-2$, and a superpotential
\begin{equation}
W \: = \: p Q(\phi),
\end{equation}
where $Q$ is a degree-two polynomial.  This theory has a mixed Coulomb-Higgs
branch.  The Coulomb vacua are solutions of
\begin{equation}  \label{eq:pn2:coulomb}
\sigma^{n+1} \: = \: (-2 \sigma)^2 q,
\end{equation}
or $\sigma^{n-1} \propto q$, which has $n-1$ solutions.  In addition,
there is a Landau-Ginzburg orbifold, a ${\mathbb Z}_2$ orbifold
of a theory with superpotential of the form
\begin{equation}
W \: = \: Q(\phi).
\end{equation}
The $\phi$ fields are massive, and there are $n+1$ of them.
If $n+1$ is odd, then taking the ${\mathbb Z}_2$ orbifold results in a
single vacuum, whereas if $n+1$ is even, then taking the
${\mathbb Z}_2$ orbifold results in a pair of vacua 
\cite{Hellerman:2006zs,Caldararu:2007tc,Hori:2011pd}.
Combining the Coulomb and Landua-Ginzburg vacua, we see that if
$n+1$ is odd, there are $n$ total vacua,
and if $n+1$ is even, there are $n+1$ total vacua,
which matches the Euler characteristic of ${\mathbb P}^n[2]$,
as expected.

As a simple example, consider ${\mathbb P}^5[2] = G(2,4)$.
The elements of the classical cohomology ring are associated to
Young tableaux
\begin{equation}
1, \: \: \:
{\tiny\yng(1)}, \: \: \:
{\tiny\yng(2)}, \: \: \: {\tiny\yng(1,1)}, \: \: \:
{\tiny\yng(2,1)}, \: \: \:
{\tiny\yng(2,2)},
\end{equation}
with degree equal to the number of boxes, and relations determined by
Schur polynomials, for example,
\begin{equation}
\left( {\tiny\yng(1)} \right)^2 \: = \: {\tiny\yng(2)} \: + \: {\tiny\yng(1,1)}.
\end{equation}
The Coulomb branch computation of the GLSM for ${\mathbb P}^5[2]$ sees
only the restriction of cohomology from the ambient ${\mathbb P}^5$.
In that language, $\sigma$ is naturally associated to $\tiny\yng(1)$, but
there is only one $\sigma^2$, whereas in fact the cohomology ring has
an additional generator in middle degree, reflected here in the fact that
both $\tiny\yng(2)$ and $\tiny\yng(1,1)$ have degree two.
The extra generator is realized physically as the
Landau-Ginzburg orbifold vacuum.  It is natural to conjecture that the
quantum symmetry of that ${\mathbb Z}_2$ Landau-Ginzburg orbifold
corresponds to transposition of the Young tableaux, so that the 
Landau-Ginzburg twist field is proportional to 
the difference 
\begin{equation}
{\tiny\yng(2)} - {\tiny\yng(1,1)}.
\end{equation}

Now, let us turn to quantum K theory rings.
From the general formula~(\ref{eq:chiralring:pn-d}) 
for ${\mathbb P}^n[d]$ for $n$ even, we have for $d=2$ the ring relation
\begin{equation}
(1 - x)^{n+1} \: = \: q \left( 1 - x^2 \right)^2,
\end{equation}
or more simply,
\begin{equation}
 \left( 1 - x \right)^{n-1}
\: = \:
 q (1 + x)^2.
\end{equation}

Mathematically, the quantum K theory ring in this case ($n$ even) is
\begin{equation}
{\mathbb C}[y][q] / \left\langle y^n - q y (y-2)^2 \right\rangle,
\end{equation}
where $y$ is the divisor class, e.g. ${\cal O}_{\tiny\yng(1)}$.
This matches the relations above if we 
identify $y = 1-x$.

In more general cases, one must take into account contributions from
the Landau-Ginzburg orbifold, as this is a mixed Higgs-Coulomb branch.
Such computations are discussed in detail in
\cite{mixedope}.

\section{Symplectic Grassmannians}
\label{sect:sg}

In this section, we will make some general remarks on physical
predictions for quantum K theory rings of symplectic Grassmannians.
Mathematically, quantum K theory rings are known for the Lagrangian
Grassmannians, but not necessarily for the other symplectic Grassmannians.
We will compare our predictions to known results for $LG(2,4)$ in a basis
of Schubert classes,
make some predictions for quantum K theory rings of 
more general symplectic Grassmannians in a shifted Wilson line basis,
and also propose the quantum K theory rings of Lagrangian Grassmannians
in terms of $\lambda_y$ classes, which we intend to address mathematically
in \cite{leom}.

In section~\ref{sect:sgk2n:exs} we will compare those
mathematical results and physical predictions, in concrete examples
of Lagrangian Grassmannians for which mathematical results are known.

\subsection{General remarks}

As discussed in \cite{Gu:2020oeb,ot}, the GLSM for a symplectic Grassmannian
$SG(k,2n)$ is a $U(k)$ gauge theory with $2n$ fundamentals $\Phi_{\pm i}$
($i \in \{1, \cdots, n\}$),
one chiral superfield $q$ in the representation $\wedge^2 V^*$ (for $V$
the fundamental), and superpotential
\begin{equation}
W \: = \: \sum_{i=1}^n \sum_{a, b} q_{ab} \Phi^a_i \Phi^b_{-i}.
\end{equation}
In the special case $k=n$, $SG(k,2n)$ is also referred to as a Lagrangian
Grassmannian, and denoted $LG(n,2n)$.

We are going to use Coulomb branch methods to analyze this theory.
Now, this assumes that the IR is described by a pure Coulomb
branch.  As discussed in \cite[section 2.1]{Gu:2020oeb}, we only have
a pure Coulomb branch when $k$ is odd.  In addition, in the special
case that $k=2=n$, although there is a Higgs branch, it only contributes
a single state, which (for reasons discussed in \cite{mixedope}, 
will not alter our computations.  Therefore, in our analysis of $SG(k,2n)$,
we will assume throughout that $k$ is either odd, or $k=2=n$.

From the general expressions in section~\ref{sect:rev},
the twisted one-loop effective superpotential for this theory
has the form
\begin{eqnarray}
W & = &
\frac{1}{2} k_{SU(k)} \sum_a \left( \ln x_a \right)^2
\: + \:
\frac{ k_{U(1)} - k_{SU(k)} }{2k} \left( \sum_a \ln x_a \right)^2
\nonumber \\
& &
\: + \: \left(\ln (-)^{k-1} q\right) \sum_a \ln x_a \: + \:
2 n \sum_a {\rm Li}_2\left(  x_a \right) 
\: + \: \sum_{a < b} {\rm Li}_2\left( x_a^{-1} x_b^{-1} \right)
\nonumber \\
& & 
\: + \: \frac{2n}{4} \sum_a \left( \ln x_a \right)^2
\: + \: \frac{1}{4} \sum_{a < b} \left( \ln x_a + \ln x_b \right)^2,
\\
& = &
\frac{1}{4} \left( 2 k_{SU(k)} + (k-2) + 2n \right)
\sum_a \left( \ln x_a \right)^2 
\: + \:
\frac{\left( 2\left( k_{U(1)} - k_{SU(k)}\right) + k \right)
}{4k}
\left( \sum_a \ln x_a \right)^2
\nonumber \\
& &  
\: + \: \left(\ln (-)^{k-1} q\right) \sum_a \ln x_a \: + \:
2 n \sum_a {\rm Li}_2\left(  x_a \right) 
\: + \: \sum_{a < b} {\rm Li}_2\left( x_a^{-1} x_b^{-1} \right).
\label{eq:lgk2n:W}
\end{eqnarray}

If there were no superpotential, if we were describing the total space
of a vector bundle on $G(k,2n)$, we would take
\begin{eqnarray}
k_{U(1)} & = & - (2n)/2 - k-1 \: = \: - n - k + 1,
\\
k_{SU(k)} & = & k - (2n)/2 - (1/2) T_2(R),
\end{eqnarray}
for $R = \wedge^2 {\bf k}$, ${\bf k}$ the fundamental of $SU(k)$.
(We take the dual of the representation appearing, as this is the contribution
from the integrated-out $N=2$ chiral that formed half of an $N=4$ 
hypermultiplet in three dimensions.)
In the expression above, we used the identity
\begin{equation}   \label{eq:identity1}
\sum_{a < b} \left( \ln x_a + \ln x_b \right)^2 \: = \:
(k-2) \sum_a \left( \ln x_a \right)^2 \: + \:
\left( \sum_a \ln x_a \right)^2.
\end{equation}
For any representation $R$ of $SU(k)$,
\begin{equation}
T_2(R) \: = \: \frac{1}{k} \frac{ \dim R }{\dim SU(k) } C_2(R),
\end{equation}
and in this case,
\begin{equation}
T_2( \wedge^2 {\bf k} ) \: = \: k-2.
\end{equation}

However, we do have a superpotential, and in this case,
following the observations in section~\ref{sect:hyp:gen},
we conjecture the following levels
\begin{eqnarray}
k_{U(1)} & = &  k - n -1,
\\
k_{SU(k)} & = & (3k - 2n - 2)/2 \: = \: (3/2) k - n - 1 ,
\end{eqnarray}
for a symplectic Grassmannian $SG(k,2n)$.  
(We will justify this choice by comparing the resulting physical
predictions against known mathematics.  That said, quantum K theory
is only known for Lagrangian Grassmannians ($k=n$), not more general
cases, so as a result, our checks on these proposed levels are
necessarily limited.)

Plugging in, we have
the superpotential
\begin{eqnarray}
W & = &
(k-1) \sum_a \left( \ln x_a \right)^2 \: + \:
\left[ \ln \left( (-)^{k-1} q \right) \right] \sum_a \ln x_a
\nonumber \\
& &
\: + \:
2n \sum_a {\rm Li}_2(x_a) \: + \: \sum_{a < b} {\rm Li}_2\left( x_a^{-1}
x_b^{-1} \right),
\label{eq:sgk2n:sup1}
\end{eqnarray}
from which we derive the physical ring relations (as derivatives of $W$):
\begin{equation}  \label{eq:sgk2n:phys}
(-)^{k-1} q \, x_a^{2k-2}  \left( \prod_{c \neq a} \left( 1 -
x_a^{-1} x_c^{-1} \right) \right)
\: = \: (1-x_a)^{2n}.
\end{equation}
In much of this section, we will focus on Lagrangian Grassmannians
$LG(n,2n)$, for which the ring relations above reduce to
\begin{equation}  \label{eq:lgn2n:phys}
(-)^{n-1} q \, x_a^{2n-2} \left( \prod_{c \neq a} \left( 1 -
x_a^{-1} x_c^{-1} \right) \right)
\: = \: (1-x_a)^{2n}.
\end{equation}

Now, in the gauge theory, the excluded loci are
\begin{equation}
\sigma_a \: \neq \: \pm \sigma_b
\end{equation}
for $a \neq b$ ($\sigma_a \neq \sigma_b$ is typical for a $U(k)$
gauge theory, and the condition $\sigma_a \neq - \sigma_b$ is
a consequence of the presence of the antisymmetric two-tensor
\cite[section 7]{Gu:2018fpm}, \cite[appendix C]{Gu:2020oeb}.
As a result, 
\begin{equation}
x_a \neq x_b^{\pm 1}
\end{equation}
for $a \neq b$, and so we can divide out factors of
$x_a - x_b$, $1 - x_a x_b$ from the equations of motion above.

In passing, let us also take a moment to compare the $R \rightarrow 0$
limit to the quantum cohomology ring of a symplectic Grassmannian,
as a consistency check.
To this end, we rewrite~(\ref{eq:sgk2n:phys}) as
\begin{equation}
(-)^{k-1} q_{3d} \, x_a^{k-1} \left( \prod_{c \neq a}\left(
x_a - x_c^{-1} \right) \right) \: = \: (1 - x_a)^{2n}.
\end{equation}
Now, for small $R$,
\begin{eqnarray}
x_a^{k-1} & \mapsto & 1,
\\
x_a - x_c^{-1} & \mapsto & R \left( \sigma_a + \sigma_c \right),
\\
1 - x_a & \mapsto & R \sigma_a,
\\
q_{3d} & \mapsto & R^{2n - (k-1)} q_{2d},
\end{eqnarray}
where we have suppressed factors of $2 \pi i$, and
in relating $q_{3d}$ to $q_{2d}$, the power of $R$ is determined
by the axial anomaly.
Plugging these in, we see that~(\ref{eq:sgk2n:phys}) becomes
\begin{equation}
(-)^{k-1} q_{2d} R^{2n} \prod_{c \neq a} \left( \sigma_c +
\sigma_a \right) \: = \: 
R^{2n} \sigma_a^{2n}.
\end{equation}
Thus, cancelling out factors of $R$, we recover 
the quantum cohomology relation for $SG(k,2n)$ given
in \cite[section 2.3]{Gu:2020oeb}, as expected.

\subsection{$LG(2,4)$ and Schubert classes}

As a consistency check (for example, on our choices of Chern-Simons
levels), in this section we will explicitly compute the physics
predictions for the quantum K theory ring and compare to 
existing mathematical results in terms of Schubert classes.
Starting from physics, expressing everything
in a Schubert basis is computationally intensive, so in the next two
sections we will 
propose a description of
the quantum K theory ring of
symplectic and Lagrangian Grassmannians in
terms of two new bases, of shifted Wilson lines and $\lambda_y$ classes.  
We will compare those new rings to existing
results in section~\ref{sect:sgk2n:exs}, and intend to address the
matter mathematically 
in \cite{leom}.

From equation~(\ref{eq:sgk2n:sup1}) we have the superpotential
\begin{eqnarray}
W & = &
\sum_a \left( \ln x_a \right)^2 
\: + \:(\ln (-q)) \sum_a \ln x_a \: + \:
2 n \sum_a {\rm Li}_2\left(  x_a \right) 
\: + \: \sum_{a < b} {\rm Li}_2\left( x_a^{-1} x_b^{-1} \right).
\end{eqnarray}
From this one derives the critical locus equations~(\ref{eq:lgn2n:phys}),
which for $LG(2,4)$ are
\begin{eqnarray}
(1 - x_1)^4 & = & - q x_1^2 \left(1 - x_1^{-1} x_2^{-1} \right),
\label{eq:lg24:A}
\\
(1 - x_2)^4 & = & - q x_2^2 \left( 1 - x_1^{-1} x_2^{-1} \right).
\label{eq:lg24:B}
\end{eqnarray}

These critical locus equations imply
\begin{equation}
x_2^2 (1 - x_1)^4 \: = \: x_1^2 (1 - x_2)^4,
\end{equation}
and after factoring out $x_1 - x_2$, $1 - x_1 x_2$, this becomes
\begin{equation}  \label{eq:lg24:crit1}
4 x_1 x_2 \: = \: (x_1 + x_2) (1 + x_1 x_2).
\end{equation}

Taking the difference of $x_2$ times (\ref{eq:lg24:A}) and $x_1$ times
(\ref{eq:lg24:B}), and factoring out $x_1-x_2$ (which is nonzero because
of the excluded locus condition), we find
\begin{equation}  \label{eq:lg24:crit1a}
-1 + 6 x_1 x_2 - 4 x_1 x_2 (x_1 + x_2) + x_1 x_2 \left(x_1^2 + x_1 x_2 + x_2^2
\right) \: = \: - q (x_1 x_2 - 1).
\end{equation}
Twice this plus
\begin{equation}
- \left( -2 + x_1 + x_2 \right) \left( x_1 + x_2 - 4 x_1 x_2
+ x_1^2 x_2 + x_1 x_2^2 \right)
\end{equation}
(which vanishes from (\ref{eq:lg24:crit1}), and factoring out $x_1 x_2 - 1$
(which is nonzero because of the excluded locus condition), we have
\begin{equation}  \label{eq:lg24:crit2}
2 - 2(x_1 + x_2) + x_1^2 + x_2^2 \: = \: - 2q.
\end{equation}

For ordinary Grassmannians, we discussed Wilson lines associated to
various representations of $SU(2)$ (or, if the reader prefers,
to the Schur functor applied to the universal subbundle $S$).
In principle, the same Wilson lines
arise here, e.g., $W_{\tiny\yng(1)} = x_1 + x_2$.  However, because of e.g.
equation~(\ref{eq:lg24:crit1}), not all the Wilson lines are independent.
In particular, equation~(\ref{eq:lg24:crit1}) implies that
\begin{equation}
4 W_{\tiny\yng(1,1)} \: = \: W_{\tiny\yng(1)} + W_{\tiny\yng(2,1)}.
\end{equation}
As a result, since they are not independent,
we will not consider either $W_{\tiny\yng(1,1)}$ or
$W_{\tiny\yng(2,2)} = W_{\tiny\yng(1,1)}^2$.  Mathematically, this corresponds
to the fact that there are fewer Schubert cycles in $LG(2,4)$ than
the ordinary Grassmannian $G(2,4)$.

Classically, the Schubert classes and (the remaining) Wilson loop operators
(Schur functors) are related mathematically by
\begin{eqnarray}
W_{\tiny\yng(1)} & = & 2 - {\cal O}_{\tiny\yng(1)} -
{\cal O}_{\tiny\yng(2)},
\\
W_{\tiny\yng(2)} & = & 3 = 3 {\cal O}_{\tiny\yng(1)} -
2 {\cal O}_{\tiny\yng(2)} + {\cal O}_{\tiny\yng(2,1)},
\\
W_{\tiny\yng(2,1)} & = & 2 - 3 {\cal O}_{\tiny\yng(1)} +
{\cal O}_{\tiny\yng(2)},
\end{eqnarray}
or equivalently,
\begin{eqnarray}
{\cal O}_{\tiny\yng(1)} & = &
1 - \frac{1}{4} W_{\tiny\yng(1)} - \frac{1}{4} W_{\tiny\yng(2,1)},
\\
{\cal O}_{\tiny\yng(2)} & = &
1 - \frac{3}{4} W_{\tiny\yng(1)} + \frac{1}{4} W_{\tiny\yng(2,1)},
\\
{\cal O}_{\tiny\yng(2,1)} & = &
2 - \frac{9}{4} W_{\tiny\yng(1)} + W_{\tiny\yng(2)}
- \frac{1}{4} W_{\tiny\yng(2,1)}.
\end{eqnarray}

For ordinary Grassmannians, those classical mathematical relationships
also hold in quantum K theory.  In the quantum K theory of Lagrangian
Grassmannians, however, these relations receive quantum corrections.
Specifically, the quantum-corrected relationship for
$LG(2,4)$ is given by
\begin{eqnarray}
        \mathcal{O}_{\tiny \yng(1)} & = & 1 - \frac{1}{4} W_{\tiny \yng(1)} - \frac{1}{4} W_{\tiny \yng(2,1)}, \\
        \mathcal{O}_{\tiny \yng(2)} & = & 1 - \frac{3}{4} W_{\tiny \yng(1)} + \frac{1}{4} W_{\tiny \yng(2,1)},\\
        \mathcal{O}_{\tiny \yng(2,1)} & = & 2 - \frac{9}{4} W_{\tiny \yng(1)} + W_{\tiny \yng(2)} - \frac{1}{4} W_{\tiny \yng(2,1)} + q.
\end{eqnarray}

Note that ${\cal O}_{\tiny\yng(2,1)}$ has a term proportional to $q$, so we
are positing that the relations between Schubert classes and Wilson loop
operators are not solely determined classically.  In fact, we expect
that this should be true in general, and that ordinary Grassmannians
represent an unusual special case.

Using the Wilson line operators (same as for $G(2,4)$)
\begin{eqnarray}
W_{\tiny\yng(1)} & = & x_1 + x_2,
\\
W_{\tiny\yng(2)} & = & x_1^2 + x_2^2 + x_1 x_2,
\\
W_{\tiny\yng(2,1)} & = & x_1^2 x_2 + x_1 x_2^2,
\end{eqnarray}
we find
\begin{eqnarray}
{\cal O}_{\tiny\yng(1)} & = & 1 \: - \: 
1 - (1/4)\left( x_1 + x_2 + x_1^2 x_2 + x_1 x_2^2 \right),
\\
& = & 1 - x_1 x_2,
\\
{\cal O}_{\tiny\yng(2)} & = &
1 - (3/4)( x_1 + x_2) + (1/4) x_1 x_2 (x_1 + x_2),
\\
& = &
1 - (x_1 + x_2) + x_1 x_2,
\\
{\cal O}_{\tiny\yng(2,1)} & = & 
2 - (9/4) (x_1 + x_2) + (x_1^2 + x_2^2 + x_1 x_2) - (1/4) x_1 x_2 (x_1 +
x_2) + q,
\\
& = & - 2 q + q \: = \: - q, 
\end{eqnarray}
where we have used equations~(\ref{eq:lg24:crit1}), (\ref{eq:lg24:crit2}).

Using equations~(\ref{eq:lg24:crit1}), (\ref{eq:lg24:crit2}),
it is straightforward to show that
\begin{eqnarray}
{\cal O}_{\tiny\yng(1)} \cdot {\cal O}_{\tiny\yng(1)}
 & = &
q - q {\cal O}_{\tiny\yng(1)} + 2 {\cal O}_{\tiny\yng(2)} -
{\cal O}_{\tiny\yng(2,1)},
\\
{\cal O}_{\tiny\yng(2)} \cdot {\cal O}_{\tiny\yng(1)}
& = &
-q + q {\cal O}_{\tiny\yng(1)} + {\cal O}_{\tiny\yng(2,1)},
\\
{\cal O}_{\tiny\yng(2,1)} \cdot {\cal O}_{\tiny\yng(1)}
& = &
- q {\cal O}_{\tiny\yng(1)},
\\
{\cal O}_{\tiny\yng(2)} \cdot {\cal O}_{\tiny\yng(2)}
& = &
- q {\cal O}_{\tiny\yng(1)},
\\
{\cal O}_{\tiny\yng(2,1)} \cdot {\cal O}_{\tiny\yng(2)}
& = &
- q {\cal O}_{\tiny\yng(2)},
\\
{\cal O}_{\tiny\yng(2,1)} \cdot {\cal O}_{\tiny\yng(2,1)}
& = & q^2.
\end{eqnarray}
which matches known mathematical results for the quantum K theory ring
\cite{buch-mih1,chap-perr,bcmp,buchcomp}.

As a minor point of interest, the reader should note that
classically, the Schubert cycle ${\cal O}_{\tiny\yng(2,1)}$ has
vanishing products with all other Schubert cycles, reflecting the
fact that ${\cal O}_{\tiny\yng(2,1)} \sim - q$.
To be clear, this does not imply that the Schubert class
${\cal O}_{\tiny\yng(2,1)}$ vanishes classically, and in fact, it does
not.  Instead, that Schubert class should be understood as
a subvariety of the Higgs branch, whereas all computations in this
section take place on the Coulomb branch, and so 
are akin to merely an index of the Higgs branch subvariety.

\subsection{Shifted Wilson line basis for symplectic Grassmannians}
\label{sect:sgk2n:sw}

In this section, we will derive from physics a shifted Wilson line
basis for the quantum K theory ring of symplectic Grassmannians $SG(k,2n)$.
Now, so far as we are aware, quantum K theory is only known in mathematics
for Lagrangian Grassmannians, corresponding to the special cases $k=n$.
Thus, in this section we will in principle be making a physical prediction for
the quantum K theory ring of general symplectic Grassmannians,
for $k$ odd or $k=2=n$ (so that pure Coulomb branch methods
apply, as discussed earlier),
and in later sections, we will check that prediction\footnote{
To be clear, our prediction for general symplectic Grassmannians
$SG(k,2n)$ will only be as good as our choice of Chern-Simons levels,
which we have only been able to check in the case of Lagrangian
Grassmannians, for which $k = n$.
} against known
results for Lagrangian Grassmannians, as well as against $\lambda_y$
class relations for Lagrangian Grassmannians which will be described next.

First, we rewrite the physical ring relation~(\ref{eq:sgk2n:phys})
in the form
\begin{equation}  \label{eq:sgk2n:sw:phys}
(-)^{k-1} q \left(1 - z_a\right)^{k-1} \prod_{b \neq a}\left(
z_a z_b - z_a - z_b \right) \: = \:
z_a^{2n} \prod_{c \neq a} \left( 1 - z_c \right),
\end{equation}
where $z_a = 1 - x_a$.

We can write
\begin{eqnarray}
\prod_{b \neq a} \left( z_a z_b - z_a - z_b \right)
 & = &
\prod_{b \neq a} \left( z_b( z_a - 1) - z_a \right),
\\
& = &
\sum_{\ell = 0}^{n-1} (-)^{n-1-\ell} z_a^{n-1-\ell} (z_a - 1)^{\ell}
e_{\ell}(z'),
\end{eqnarray}
where $z'$ denotes the collection of $z_b \neq z_a$.

Next, we use the following identities:
\begin{eqnarray}
e_i(z') & = & \sum_{j=0}^i (-)^j z_a^j e_{i-j}(z),
\: \: \:
1 \leq i \leq n-1,
\\
z_a e_i(z') & = & e_{i+1}(z) - e_{i+1}(z'),
\\
z_2^j e_i(z') & = & z_a^{j-1} e_{i+1}(z) - z_a^{j-2} e_{j+2}(z) 
+ \cdots +
(-)^{j-1} e_{i+j}(z) \: + \: 
(-)^j e_{i+j}(z'),
\end{eqnarray}
where the last identity is established inductively.

With these identities, we can write
\begin{eqnarray}
\lefteqn{
\prod_{b \neq a} \left( z_a z_b - z_a - z_b \right)
} \nonumber \\
& = &
 \sum_{p=1}^{k-1}z_a^{k-1-p} \left[ 
\sum_{i=1}^{k-1} \sum_{m=\max\{p+1-i,1\}}^{\min\{p,k-i\}} (-)^{k-i+p}  
\binom{i}{p-m} e_{i+m}(z) +  \sum_{i=p}^{k-1} (-)^{k-1-i+p} e_{p}(z) \right]
\nonumber \\
& & 
\: + \: z_a^{k-1} (1/2) \left( (-)^{k-1}+1 \right).
\end{eqnarray}
In particular, the left-hand-side of the physical chiral ring
relation~(\ref{eq:sgk2n:sw:phys}) can be written
\begin{eqnarray}
\lefteqn{
(-)^{k-1} q \left(1 - z_a\right)^{k-1} \prod_{b \neq a}\left(
z_a z_b - z_a - z_b \right) 
} \nonumber \\
& = &
(-)^{k-1} q 
\sum_{\ell=1}^{2k-2} z_a^{2k-2-\ell}  \Biggl[ 
\sum_{p=1}^{k-1} (-)^{\ell - p} \binom{k-1}{\ell - p}  
\Bigg(  \sum_{i=1}^{k-1} \sum_{m=\max\{p+1-i,1\}}^{\min\{p,k-i\}} (-)^{k-i+p}  
\binom{i}{p-m} e_{i+m}(z)
\nonumber \\
& & \hspace*{3.25in}
\: + \: \sum_{i=p}^{k-1} (-)^{k-1-i+p} e_{p}(z) \Biggr) 
\Biggr] 
\nonumber \\
& & 
   \: + \: (-)^{k-1} q (1/2) \left( (-)^{k-1}+1\right)
 \sum_{\ell=0}^{k-1} (-)^{\ell} \binom{k-1}{\ell} z_a^{2k-2-\ell}.
\end{eqnarray}

For the right-hand-side of the physical chiral ring
relation~(\ref{eq:sgk2n:sw:phys}), we can use
equation~(\ref{eq:gkn:sw:piece1}), namely
\begin{eqnarray}
\lefteqn{
z_a^n \prod_{c \neq a} \left(1 - z_c \right)
} \nonumber \\
& \hspace*{0.25in} = & z_a^n \: - \: z_a^{n-1} \left( e_2(z) - e_3(z) + e_4(z) - \cdots
+ (-)^{n-2} e_n(z) \right)
\nonumber \\
& & \hspace*{0.3in}
\: + \:
z_a^{n-2} \left( e_3(z) - e_4(z) + e_5(z) - \cdots + (-)^{n-3} e_n(z)
\right)
\nonumber \\
& & \hspace*{0.3in}
\: - \:
z_a^{n-3} \left( e_4(z) - e_5(z) + e_6(z) - \cdots + (-)^{n-4} e_n(z) \right)
\nonumber \\ 
& &  \hspace*{0.7in} 
\: + \: \cdots \: - \: 
 z_a e_n(z). 
\\
& = &
z_a^n \: - \: \sum_{p=1}^{n-1} (-)^p z_a^{n-p} \sum_{\ell=1}^{k-1}
(-)^{\ell} e_{p+\ell}(z),
\end{eqnarray}
in conventions in which $e_p(z) = 0$ for
$p < 0$ or $p > k$.

Putting this together, and replacing $z_a$ with the indeterminate
$t$, we get a characteristic polynomial associated with this
theory:
\begin{eqnarray}
\lefteqn{
t^{2n} - \sum_{p=1}^{k-1}(-)^p  t^{2n-p} \sum_{\ell = 1}^{k-1} 
(-)^{\ell} e_{p+\ell}(z)  +
(-)^{k} q (1/2) \left( 1 - (-)^{k} \right)
 \sum_{\ell=0}^{k-1} (-)^{\ell} \binom{k-1}{\ell} t^{2k-2-\ell}
} \nonumber \\
& & 
   + (-)^{k} q \sum_{\ell=1}^{2k-2} t^{2k-2-\ell} \Biggl[ 
\sum_{p=1}^{k-1} (-)^{\ell - p} \binom{k-1}{\ell - p}  \Biggl( 
 \sum_{i=1}^{k-1} \sum_{m=\max\{p+1-i,1\}}^{\min\{p,k-i\}} (-)^{k-i+p}  \binom{i}{p-m} e_{i+m}(z)
\nonumber \\
& & \hspace*{3in} 
 +  \sum_{i=p}^{k-1} (-)^{k-1-i+p} e_{p}(z) \Biggr) \Biggr]  = 0.
\label{eq:sgk2n:vieta1}
\end{eqnarray}

Next, we write this as
\begin{equation}
\prod_{\ell} \left( t - w_{\ell} \right) \: = \: 0,
\end{equation}
where the $w_{\ell}$ are all $2n$ roots of the characteristic
polynomial~(\ref{eq:sgk2n:vieta1}) (of which $k < 2n$ correspond to the
$z_a$), and comparing coefficients (i.e. using Vieta's formula), we find
\begin{eqnarray}
e_m(w) & = & \sum_{\ell = 1}^{k-1} (-)^{\ell+1} e_{m+\ell}(z) + (-)^k q 
(1/2) \left( 1 - (-)^k \right)  \binom{k-1}{m-2-2n+2k} 
\nonumber \\
& &
 + (-)^k q  \left[
\sum_{p=1}^{k-1} (-)^{p} \binom{k-1}{m-2-2n+2k - p}  \left(
 \sum_{i=1}^{k-1} \sum_{m=\max\{p+1-i,1\}}^{\min\{p,k-i\}} (-)^{k-i+p} 
 \binom{i}{p-m} e_{i+m}(z) \right. \right. 
\nonumber \\
&  & \hspace*{3in}
 \left. \left. +  \sum_{i=p}^{k-1} (-)^{k-1-i+p} e_{p}(z)  \right) \right],
\end{eqnarray}
in conventions
\begin{align*}
        e_{p}(z) = 0, &\quad \text{for $p< 0$  or $p>k$},\\
        \binom{k}{\ell}=0, &\quad \text{for $\ell < 0$  or $\ell>k$}.
\end{align*}

Proceeding as for shifted Wilson line bases for ordinary
Grassmannians in section~\ref{sect:gkn:shifted},
and writing $e_m(w)$ in terms of $e_i(z)$ and $e_j(v)$ for another
$2n-k$ variables $v$, we can use the first $2n-k$ equations above
to solve for the $e_j(v)$ in terms of the $e_i(z)$, and then the remaining
equations are constraints on the $e_i(z)$.

To perform consistency checks, 
we will compare against known results for Lagrangian Grassmannians $LG(n,2n)$.
To that end, below we give the specialization of the general formulas above
to Lagrangian Grassmannians.

For the Lagrangian Grassmannian $LG(n,2n)$, a characteristic polynomial is
\begin{eqnarray}
\lefteqn{
t^{2n} - \sum_{p=1}^{n-1}(-)^p t^{2n-p} \sum_{\ell = 1}^{n-1} (-)^{\ell} e_{p+\ell}(z)
 + (-)^n q (1/2) \left( 1 - (-)^n \right)  \sum_{\ell=0}^{n-1} (-)^{\ell} \binom{n-1}{\ell} t^{2n-2-\ell}
} \nonumber \\
& &
+ (-)^n q \sum_{\ell=1}^{2n-2} (-)^{\ell} t^{2n-2-\ell} \left[
\sum_{p=1}^{n-1} (-)^{ p} \binom{n-1}{\ell - p}  \left(
 \sum_{i=1}^{n-1} \sum_{m=\max\{p+1-i,1\}}^{\min\{p,n-i\}} (-)^{n-i+p}  \binom{i}{p-m} e_{i+m}(z)
\right. \right.
\nonumber \\
& & \hspace*{3.25in}  \left. \left.
 +  \sum_{i=p}^{n-1} (-)^{n-1-i+p} e_{p}(z)  \right) \right] =0,
\end{eqnarray}
and applying Vieta's formula as before implies
\begin{eqnarray}
e_m(w) & = &
\sum_{\ell = 1}^{n-1} (-)^{\ell+1} e_{m+\ell}(z) 
+ (-)^n q (1/2) \left( 1 - (-)^n \right)  \binom{n-1}{m-2} 
\nonumber \\
&  &
 + (-)^n q \left[\sum_{p=1}^{n-1} (-)^{ p} \binom{n-1}{m - 2 - p}  \left(
 \sum_{i=1}^{n-1} \sum_{m=\max\{p+1-i,1\}}^{\min\{p,n-i\}} (-)^{n-i+p}
  \binom{i}{p-m} e_{i+m}(z)
\right. \right.
\nonumber \\
& &  \hspace*{3in}  \left. \left.
 +  \sum_{i=p}^{n-1} (-)^{n-1-i+p} e_{p}(z)  \right) \right],
\end{eqnarray}
where the $w_{\ell}$ are the $2n$ roots of the characteristic polynomial,
which include the $n$ $z_a$,
in the conventions
\begin{align*}
        e_{p}(z) = 0, &\quad \text{for $p< 0$  or $p>n$},\\
        \binom{n}{\ell}=0, &\quad \text{for $\ell < 0$  or $\ell>n$}.
\end{align*}

As before, in principle one can write the elementary symmetric
polynomials $e_m(w)$ in terms of
elementary symmetric polynomials in the $n$ $z_a$ and another $n$ $v$'s,
use the first $n$ equations above to solve for the $e_i(v)$'s in terms of
the $e_j(z)$'s, and then use the remaining equations to constrain the values
of $e_j(z)$.  We will see this explicitly in examples later, but for the
moment, for later use, we write out the first few values of $e_m(w)$
explicitly:
\begin{itemize}
        \item $m=1$
        \begin{equation}
                e_1(w) \: = \: \sum_{\ell = 1}^{n-1} (-)^{\ell+1} e_{1+\ell}(z),
        \end{equation}
        \item $m=2$
        \begin{equation}
                e_2(w) \: = \: \sum_{\ell = 1}^{n-1} (-)^{\ell+1} e_{2+\ell}(z)
+ (-)^n q (1/2) \left( 1 - (-)^n \right),
        \end{equation}
        \item  $3 \leq m\leq n-1 $
        \begin{eqnarray}
                e_m(w) & = & 
\sum_{\ell = 1}^{n-1} (-)^{\ell+1} e_{m+\ell}(z)
 + (-)^n q (1/2) \left( 1 - (-)^n \right) \binom{n-1}{m-2} 
\nonumber  \\
  &  &
 + (-)^n q \left[\sum_{p=1}^{n-1} (-)^{ p} \binom{n-1}{m - 2 - p}  \left(
 \sum_{i=1}^{n-1} \sum_{m=\max\{p+1-i,1\}}^{\min\{p,n-i\}} (-)^{n-i+p}  \binom{i}{p-m} e_{i+m}(z)
\right. \right. \nonumber \\
& & \hspace*{3in} \left. \left. 
 +  \sum_{i=p}^{n-1} (-)^{n-1-i+p} e_{p}(z)  \right) \right],
        \end{eqnarray}
        \item $m = n, n+1$
        \begin{eqnarray}
                e_m(w) & = &
 (-)^n q (1/2) \left( 1 - (-)^n \right)  \binom{n-1}{m-2} 
\nonumber \\
 &  &
 + (-)^n q \left[\sum_{p=1}^{n-1} (-)^{ p} \binom{n-1}{m - 2 - p}  \left(
 \sum_{i=1}^{n-1} \sum_{m=\max\{p+1-i,1\}}^{\min\{p,n-i\}} (-)^{n-i+p}  \binom{i}{p-m} e_{i+m}(z)
\right. \right. \nonumber \\
& & \hspace*{3in} \left. \left.
 +  \sum_{i=p}^{n-1} (-)^{n-1-i+p} e_{p}(z)  \right) \right],
        \end{eqnarray}
        \item $m > n+1$
        \begin{eqnarray}
                e_m(w) & = &
 (-)^n q \left[\sum_{p=1}^{n-1} (-)^{ p} \binom{n-1}{m - 2 - p}  \left(
 \sum_{i=1}^{n-1} \sum_{m=\max\{p+1-i,1\}}^{\min\{p,n-i\}} (-)^{n-i+p}  \binom{i}{p-m} e_{i+m}(z)
\right. \right. \nonumber \\
& & \hspace*{3in} \left. \left.
 +  \sum_{i=p}^{n-1} (-)^{n-1-i+p} e_{p}(z)  \right) \right],
        \end{eqnarray}
\end{itemize}

Finally, we should elaborate on how this shifted Wilson line basis
is related to e.g. Schubert classes, much as we did for
ordinary Grassmannians.  We provide the dictionary in
some special cases below.  This dictionary was derived 
using the program ``Equivariant Schubert Calculator" \cite{buchcomp}
to compute multiplications in the classical and quantum K theory rings.

For $LG(2,4)$, we compute that the dictionary is
\begin{eqnarray}
e_i(z) & = & SW_{1^i},
\\
e_1(v) & = & - {\cal O}_{\tiny\yng(1)},
\\
e_2(v) & = & {\cal O}_{\tiny\yng(2)}.
\end{eqnarray}
We will confirm this dictionary explicitly later in section~\ref{sect:ex:lg24}.

For $LG(3,6)$, we compute that the dictionary is
\begin{eqnarray}
e_i(z) & = & SW_{1^i},
\\
e_1(v) & = & - {\cal O}_{\tiny\yng(1)},
\\
e_2(v) & = & {\cal O}_{\tiny\yng(2)} - q,
\\
e_3(v) & = & - {\cal O}_{\tiny\yng(3)} - q.
\end{eqnarray}
We will discuss this example explicitly later in
section~\ref{sect:ex:lg36}.

For $LG(4,8)$, we compute that the dictionary is
\begin{eqnarray}
e_i(z) & = & SW_{1^i},
\\
e_1(v) & = & - {\cal O}_{\tiny\yng(1)},
\\
e_2(v) & = & {\cal O}_{\tiny\yng(2)},
\\
e_3(v) & = & - {\cal O}_{\tiny\yng(3)} + q,
\\
e_4(v) & = & {\cal O}_4 + q.
\end{eqnarray}

For $LG(5,10)$, we compute that the dictionary is
\begin{eqnarray}
e_i(z) & = & SW_{1^i},
\\
e_1(v) & = & - {\cal O}_{\tiny\yng(1)},
\\
e_2(v) & = & {\cal O}_{\tiny\yng(2)} - q,
\\
e_3(v) & = & - {\cal O}_{\tiny\yng(3)} - 3q,
\\
e_4(v) & = & {\cal O}_4 - 4q,
\\
e_5(v) & = & - {\cal O}_5 - 2q.
\end{eqnarray}

We shall study in detail how these are applied in examples later.
Next, however, we will present a $\lambda_y$-class description of the quantum
K theory ring relations for Lagrangian Grassmannians.

\subsection{$\lambda_y$ class relations for $LG(n,2n)$}
\label{sect:lambda:lgn2n}

In this section we will propose a description of  
the quantum K theory ring of $LG(n,2n)$ in terms of the
$\lambda_y$ class of the universal subbundle, much as we
did for ordinary Grassmannians in section~\ref{sect:lambda:gkn}.
We refer the reader to that section for notation.

On the ambient ordinary Grassmannian $G(n,2n)$, there is the canonical
exact sequence
\begin{equation}
0 \: \longrightarrow \: S \: \longrightarrow \:
{\cal O}^{2n} \: \longrightarrow \: Q \: \longrightarrow \: 0,
\end{equation}
which one can restrict to $LG(n,2n)$.  
In the rest of this section,
we will use $S$, $Q$ to denote the restrictions of $S$ and $Q$ on
the ambient $G(n,2n)$ to $LG(n,2n)$.  Along the restriction to
$LG(n,2n)$, $Q \cong S^*$.
Then, in terms of $\lambda_y$ classes, we have classically
that
\begin{equation}
\lambda_y(S) \lambda_y(Q) \: = \: (1+y)^{2n},
\end{equation}
where $y$ is a formal variable.

We propose 
that quantum corrections can be incorporated by writing
\begin{equation}  \label{eq:lambda-q}
\lambda_y(S) \star \lambda_y(Q) \: = \: (1+y)^{2n} \: + \: F(y,q),
\end{equation}
where\footnote{
In mathematics conventions for Lagrangian Grassmannians, the $q$ differs
from that used here by a sign,
\begin{equation}
q_{\rm math} \: = \: (-)^{n-1} q_{\rm phys} \: = \:  (-)^{n-1} q.
\end{equation}
For Grassmannians, there is no difference in conventions.
}
\begin{equation}
F(y,q) \: = \: - (-)^{n-1} q \sum_{i=0}^{2n} R_i y^i,
\end{equation}
with the $R_i$ determined as follows:
\begin{itemize}
\item $R_0 = 0$, $R_1 = 0$,
\item $R_i = R_{2n+2-i}$ for all $2 \leq i \leq n+1$,
\item For any $2 \leq i \leq n+1$,
\begin{equation}
R_i \: = \: \wedge^{i-2} Q + \wedge^{i-4} Q + \cdots +
\wedge^{i - 2[i/2]} Q,
\end{equation}
in conventions in which $\wedge^0 S^*  = 1$ and $\wedge^{-i} S^* = 0$.
\end{itemize}
(We intend to address this mathematically in \cite{leom}.)

For $LG(2,4)$, we have $R_2 = R_4 = 1$, $R_3 = Q$, hence
\begin{equation}
F(y,q) \: = \: + q y^2( 1 + y Q + y^2). 
\end{equation}
Formally writing, from the splitting principle, $S = \oplus_a x_a$,
$Q = \oplus_a \tilde{x}_a$,
we have
\begin{eqnarray}
\lambda_y(S) & = & 1 + y(x_1 + x_2) + y^2(x_1 x_2) \: = \:
1 + y e_1(x) + y^2 e_2(x),
\\
\lambda_y(Q) & = & 1 + y( \tilde{x}_1 + \tilde{x}_2) + y^2 (\tilde{x}_1
\tilde{x}_2) \: = \: 1 + y e_1(\tilde{x}) + y^2 e_2(\tilde{x}),
\end{eqnarray}
where $e_i$ denote
the elementary symmetric polynomials.
As a result,
the coefficients of powers of $y$ in equation~(\ref{eq:lambda-q}) are
\begin{eqnarray}
e_1(x) + e_1(\tilde{x}) & = & 4,
\\
e_2(x) + e_1(x) e_1(\tilde{x}) + e_2(\tilde{x}) & = &
6 + q,
\\
e_2(x) e_1(\tilde{x}) + e_1(\tilde{x}) e_2(x) & = &
4 + q e_1(\tilde{x}),
\\
e_2(x) e_2(\tilde{x}) & = &
1 + q.
\end{eqnarray}
We will explicitly verify that these predictions match physics and
existing mathematics in section~\ref{sect:ex:lg24:lambda}.

For $LG(3,6)$, we have $R_2 = R_6 = 1$, $R_3 = R_5 = Q$,
$R_4 = \wedge^2 Q + 1$, hence
\begin{equation}    \label{eq:lg:ex36intro}
F(y,q) \: = \: - q y^2 \left( 1 + y Q + y^2 (1 + \wedge^2 Q)
+ y^3 Q + y^4 \right).
\end{equation}
We will compare this to physics predictions and to
existing mathematics in section~\ref{sect:ex:lg36:lambda}.

For $LG(4,8)$, we have $R_2 = R_8 = 1$, $R_3 = R_7 = Q$,
$R_4 = R_6 = \wedge^2 Q + 1$, $R_5 = \wedge^3 Q + Q$, hence
\begin{equation}
F(y,q) \: = \: + q y^2\left(
1 + y Q + y^2 ( 1 + \wedge^2 Q) + y^3 ( Q + \wedge^3 Q) +
y^4 (1 + \wedge^2 Q) + y^5 Q + y^6 \right).
\end{equation}

Next, we will apply these bases to specific examples.

\section{Examples of Lagrangian Grassmannians}
\label{sect:sgk2n:exs}

In this section we will check the descriptions of quantum K theory given
in the previous section, in terms of shifted Wilson lines and
$\lambda_y$ classes, in some specific examples of
Lagrangian Grassmannians.

\subsection{$LG(2,4)$}
\label{sect:ex:lg24}

In this section we will 
describe the quantum K theory ring of $LG(2,4)$ in bases of shifted Wilson
lines and $\lambda_y$ classes, comparing to existing results as a 
consistency check.

\subsubsection{Shifted Wilson line basis}

In terms of the variables $z_a = 1 - x_a$,
the critical locus equations for $LG(2,4)$ are~(\ref{eq:lgn2n:phys}), 
\begin{equation}
- q (1-z_a) \prod_{b \neq a} \left( z_b z_a - z_a - z_b \right)
\: = \: z_a^4 \prod_{b \neq a} ( 1 - z_b),
\end{equation}
which imply a characteristic polynomial, as in
section~\ref{sect:sgk2n:sw},
\begin{equation}
t^4 - t^3 e_2(z)   - t q \left( e_2(z) - e_1(z) \right)
 + q\left(  e_2(z) - e_1(z) \right)
\: = \: 0.
\end{equation}

Letting $w_{\ell}$ denote the roots of this polynomial, and matching against
\begin{equation}
\prod_{\ell} \left( t - w_{\ell} \right) \: = \: 0,
\end{equation}
we find 
\begin{eqnarray}
e_1(w) & = & e_2(z),
\\
e_2(w) & = & 0,
\\
e_3(w) & = & q \left( e_2(z) - e_1(z) \right),
\\
e_4(w) & = & q \left(  e_2(z) - e_1(z) \right).
\end{eqnarray}
Letting $v$ denote the two roots which are different from the $z_a$, we have
\begin{eqnarray}
e_1(w) & = & e_1(z) + e_1(v),
\\
e_2(w) & = & e_2(z) + e_1(z) e_1(v) + e_2(v),
\\
e_3(w) & = &  e_2(z) e_1(v) + e_1(z) e_2(v),
\\
e_4(w) & = & e_2(z) e_2(v),
\end{eqnarray}
hence
\begin{eqnarray}
e_1(v) & = & e_2(z) - e_1(z),
\\
e_2(v) & = &  - e_2(z) - e_1(z)\left( e_2(z) - e_1(z) \right),
\end{eqnarray}
and the constraint equations
\begin{eqnarray}
e_2(z)^2 - 2 e_1(z) e_2(z) - e_1(z)^2 e_2(z) + e_1(z)^3 & = &
q \left( e_2(z) - e_1(z) \right),   \label{eq:lg24:vieta:constr1}
\\
- e_2(z)^2 - e_1(z) e_2(z)^2 + e_1(z)^2 e_2(z) & = &
q \left( e_2(z) - e_1(z) \right).  \label{eq:lg24:vieta:constr2}
\end{eqnarray}
Combining these we get the $q$-independent expression
\begin{equation}
2 e_2(z)^2 - 2 e_1(z) e_2(z) - 2 e_1(z)^2 e_2(z) + e_1(z)^3 + e_1(z) e_2(z)^2
\: = \: 0.
\end{equation}

Using the identities in section~\ref{sect:useful-identities}, 
in $x_a$ variables these
expressions are
\begin{equation}   \label{eq:lg24:vieta:1}
1 - 6 e_2(x) + 4 e_1(x) e_2(x) - e_1(x)^2 e_2(x) + e_2(x)^2 \: = \:
q \left( e_2(x) - 1 \right),
\end{equation}
\begin{equation}  \label{eq:lg24:vieta:2}
- \left( e_1(x) - e_2(x) - 1 \right) \left(
1 - 3 e_2(x) + e_1(x) e_2(x)
\right)
\: = \:
q \left(  e_2(x) - 1 \right),
\end{equation}
\begin{equation}  \label{eq:lg24:vieta:3}
e_1(x) - 4 e_2(x) + e_1(x) e_2(x) 
\: = \: 0,
\end{equation}
where we have used the excluded locus condition
\begin{eqnarray}
e_2(x) - 1 & \neq & 0
\end{eqnarray}
to remove factors in simplifying the above.
Equation~(\ref{eq:lg24:vieta:1}) is identical to 
equation~(\ref{eq:lg24:crit1a}) derived in the previous
subsection, and
equation~(\ref{eq:lg24:vieta:3})
is identical to equation~(\ref{eq:lg24:crit1}).
We used those two equations previously to derive all the Schubert products
from physics.  The remaining equation above, (\ref{eq:lg24:vieta:2}),
is not independent, but instead is determined by the other two.

Thus, in a nutshell, we have confirmed that the results produced by
the algorithm above do indeed match the physics predictions used earlier,
as expected.

Furthermore, in the $x$ variables, it is straightforward to show that
\begin{eqnarray}
e_1(v) & = & e_2(x) - 1 \: = \: - {\cal O}_{\tiny\yng(1)},
\\
e_2(v) & = &  1 - 3 e_2(x) + e_1(x) e_2(x),
\\
& = & 1 - e_1(x) + e_2(x) \: = \: {\cal O}_{\tiny\yng(2)},
\end{eqnarray}
where in the last line we have used the identity~(\ref{eq:lg24:vieta:3}).
This confirms the general statement made earlier in
section~\ref{sect:sgk2n:sw}.

\subsubsection{$\lambda_y$ classes}
\label{sect:ex:lg24:lambda}

In section~\ref{sect:lambda:lgn2n} the quantum K theory relations for 
$LG(2,4)$ are given 
from the $y$ coefficients of\footnote{
As observed previously, physics and mathematics conventions for
$LG(2,4)$ (but not $LG(3,6)$ or ordinary Grassmannians) differ on
$q$, so that in this case, $q_{\rm math} = - q_{\rm phys} = - q$.
}
\begin{equation}
\lambda_y(S) \star \lambda_y(Q) \: = \: (1+y)^4 \: + \: q y^2( 1 + y Q + y^2)
\end{equation}
as
\begin{eqnarray}
e_1(x) + e_1(\tilde{x}) & = & 4,  \label{eq:lg24:lambda:1}
\\
e_2(x) + e_1(x) e_1(\tilde{x}) + e_2(\tilde{x}) & = &
6 + q,  \label{eq:lg24:lambda:2}
\\
e_2(x) e_1(\tilde{x}) + e_1(x) e_2(\tilde{x}) & = &
4 + q e_1(\tilde{x}),  \label{eq:lg24:lambda:3}
\\
e_2(x) e_2(\tilde{x}) & = &
1 + q.    \label{eq:lg24:lambda:4}
\end{eqnarray}
where the $e_i$ denote elementary symmetric polynomials
in the splitting principle factors $S = \oplus_a x_a$, $Q = \oplus_a 
\tilde{x}_a$, for $S$ and $Q$ the restrictions to $LG(2,4)$ of the
universal subbundle and universal quotient bundle, respectively,
on the ambient $G(2,4)$.

In this section we will argue that these are equivalent to the physical
ring relations of the previous subsection.

We can write
\begin{equation}
e_1(x) \: = \: x_1 + x_2, \: \: \:
e_2(x) \: = \: x_1 x_2,
\end{equation}
where each $x_a = \exp(2 \pi i R \sigma_a)$.  
Similarly, since for $LG(n,2n)$, classically $Q = S^*$, we can take each
$\tilde{x}_a = x_a^{-1}$, and write
\begin{equation}
e_1(\tilde{x}) \: = \: \frac{1}{x_1} + \frac{1}{x_2}.
\end{equation}
(Note that because of equation~(\ref{eq:lg24:lambda:4}),
$e_2(\tilde{x})$ is more subtle.)
Then, the $y$ coefficient, equation~(\ref{eq:lg24:lambda:1}), 
can be written as
\begin{equation}
x_1 + x_2 + \frac{1}{x_1} + \frac{1}{x_2} \: = \: 4,
\end{equation}
or more simply,
\begin{equation}
x_1 + x_2 + x_1^2 x_2 + x_1 x_2^2 \: = \: 4 x_1 x_2,
\end{equation}
which immediately coincides with equation~(\ref{eq:lg24:crit1}),
one of the two equations of motion we derived physically from the
twisted one-loop superpotential.
For later use, in $z$ variables, this is
\begin{equation} \label{eq:lg24:lambda:class-z}
 e_1(z)^2 - 2 e_2(z) - e_1(z) e_2(z) \: = \: 0.
\end{equation}

Solving the constraint equations~(\ref{eq:lg24:lambda:1}),
(\ref{eq:lg24:lambda:2}), we have
\begin{eqnarray}
e_1(\tilde{x}) & = & 4 - e_1(x),
\\
e_2(\tilde{x}) & = & 6 + q - e_2(x) - e_1(x) \left( 4 - e_1(x)
\right),
\end{eqnarray}
so the remaining two equations~(\ref{eq:lg24:lambda:3}),
(\ref{eq:lg24:lambda:4}) become
\begin{eqnarray}
- 4 e_1(x)^2 + e_1(x)^3 + 4 e_2(x) + 6 e_1(x) - 2 e_1(x) e_2(x)
& = & 4 + q\left( 4 - 2 e_1(x) \right),
\\
e_2(x) \left( 6 - 4 e_1(x) + e_1(x)^2 - e_2(x) \right)
& = & 1 + q \left( 1 - e_2(x) \right).
\end{eqnarray}
In terms of shifted Wilson line $z$ variables, these can be written
\begin{eqnarray}
 - e_1(z)^3 + 2 e_1(z) e_2(z) & = &  + 2 q e_1(z),  \label{eq:lg24:lambda:z:1}
\\
- e_1(z)^3 + 2 e_1(z) e_2(z) + e_1(z)^2 e_2(z) - e_2(z)^2
& = &
- q\left( e_2(z) - e_1(z) \right).  \label{eq:lg24:lambda:z:2}
\end{eqnarray}
Equation~(\ref{eq:lg24:lambda:z:2}) matches
equation~(\ref{eq:lg24:vieta:constr1}), which was derived from physics.

Equation~(\ref{eq:lg24:lambda:z:1}) can be obtained from
equations~(\ref{eq:lg24:lambda:z:2}) and (\ref{eq:lg24:lambda:class-z})
as the combination
\begin{equation}
\left( e_1(z) + 2 \right) (\ref{eq:lg24:lambda:z:2}) \: + \:
\left( e_1(z) + e_1(z)^2 - e_2(z) + q \right) (\ref{eq:lg24:lambda:class-z}).
\end{equation}
Since both equations~(\ref{eq:lg24:lambda:z:2}) and (\ref{eq:lg24:lambda:class-z})
are the same as earlier physics predictions, we see that so too is
the remaining equation~(\ref{eq:lg24:lambda:z:1}), and so the
$\lambda_y$ class predictions are in agreement with the physics
predictions for the quantum K theory ring of $LG(2,4)$.

\subsection{$LG(3,6)$}
\label{sect:ex:lg36}

\subsubsection{Shifted Wilson line basis}

From equation~(\ref{eq:lgn2n:phys}), 
the ring relations predicted by physics for $LG(3,6)$ are
\begin{equation}  \label{eq:lg36:phys}
q x_a^{4} \left( \prod_{c \neq a} \left( 1 - x_a^{-1} x_b^{-1} \right)
\right) \: = \: \left( 1 - x_a \right)^6.
\end{equation}
In a basis of shifted Wilson lines $z_a = 1 - x_a$, 
using the same algebraic tricks as described in 
section~\ref{sect:gkn:shifted},
this can be written as
\begin{eqnarray}
\lefteqn{
q (1 - z_a)^2 \left( z_a^2 - z_a \left( e_2(z) - e_3(z) \right) 
+ \left( e_2(z) - e_3(z) \right) \right)
} \nonumber \\
& \hspace*{1.5in} = &
z_a^6 - z_a^5 \left( e_2(z) - e_3(z) \right) +
z_a^4 e_3(z),
\end{eqnarray}
hence the $z_a$ are three of the roots of the polynomial
\begin{eqnarray}
\lefteqn{
q (1 - t)^2 \left( t^2 - t \left( e_2(z) - e_3(z) \right) 
+ \left( e_2(z) - e_3(z) \right) \right)
} \nonumber \\
& \hspace*{1.5in} = &
t^6 - t^5 \left( e_2(z) - e_3(z) \right) +
t^4 e_3(z),
\end{eqnarray}

Writing this polynomial as
\begin{equation}
\prod_{\ell} \left( t - w_{\ell} \right)
\: = \:
t^6 \: - \: e_1(w) t^5 \: + \: e_2(w) t^4 \: + \: \cdots \: + \:
e_6(w)
\end{equation}
and comparing coefficients, we find
\begin{eqnarray}
e_1(w) & = & e_2(z) - e_3(z),
\\
e_2(w) & = & e_3(z) - q,
\\
e_3(w) & = & - q \left( e_2(z) - e_3(z) + 2 \right),
\\
e_4(w) & = & - q \left( 3 (e_2(z) - e_3(z) ) + 1 \right),
\\
e_5(w) & = & - 3 q \left( e_2(z) - e_3(z) \right),
\\
e_6(w) & = & - q \left( e_2(z) - e_3(z) \right).
\end{eqnarray}
Letting the first three roots $w_{\ell}$ be the $z_a$,
and letting the remaining three by denoted $v_{1,2,3}$,
we have the relations
\begin{eqnarray}
e_1(w) & = & e_1(z) + e_1(v),
\\
e_2(w) & = & e_2(z) + e_1(z) e_1(v) + e_2(v),
\\
e_3(w) & = & e_3(z) + e_2(z) e_1(v) + e_1(z) e_2(v) + e_3(v),
\\
e_4(w) & = & e_3(z) e_1(v) + e_2(z) e_2(v) + e_1(z) e_3(v),
\label{eq:lg36:w4}
\\
e_5(w) & = & e_3(z) e_2(v) + e_2(z) e_3(v),
\label{eq:lg36:w5}
\\
e_6(w) & = & e_3(z) e_3(v).
\label{eq:lg36:w6}
\end{eqnarray}
We can use the first three relations to solve for
$e_{1,2,3}(v)$:
\begin{eqnarray}
e_1(v) & = & e_2(z) - e_3(z) - e_1(z), 
\\
e_2(v) & = & e_3(z) - e_2(z) - e_1(z) e_2(z) + e_1(z) e_3(z) + e_1(z)^2 - q,
\\
e_3(v) & = & - e_3(z) - e_2(z)^2 + e_2(z) e_3(z) + 2 e_1(z) e_2(z) 
- e_1(z) e_3(z)
\nonumber \\
& &
 + e_1(z)^2 e_2(z) - e_1(z)^2 e_3(z) - e_1(z)^3 
\nonumber \\
& & 
- q \left( e_2(z) - e_3(z) - e_1(z) + 2 \right),
\end{eqnarray}

Plugging back in, the remaining three 
equations~(\ref{eq:lg36:w4})-(\ref{eq:lg36:w6}) become
\begin{eqnarray}
\lefteqn{
 e_1^4 - 3 e_1^2 e_2 - e_1^3 e_2 + e_2^2 + 2 e_1 e_2^2 + 2 e_1 e_3 + e_1^2 e_3 + e_1^3 e_3 - 2 e_2 e_3 - 2 e_1 e_2 e_3 + e_3^2
} \nonumber \\
& \hspace*{1.5in}= &
 q \left(1 - 2 e_1 + e_1^2 + 2 e_2 - e_1 e_2 - 3 e_3 + e_1 e_3\right),
\label{eq:lg36:physpred:1}
\end{eqnarray}
\begin{eqnarray}
\lefteqn{
 e_1^3 e_2 - 2 e_1 e_2^2 - e_1^2 e_2^2 + e_2^3 - e_1^2 e_3 + 2 e_2 e_3 + 2 e_1 e_2 e_3 + e_1^2 e_2 e_3 - e_2^2 e_3 - e_3^2 - e_1 e_3^2 
} \nonumber \\
& \hspace*{2.0in} = &
 q \left(e_2 + e_1 e_2  - e_2^2  - 4 e_3 + e_2 e_3 \right),
\end{eqnarray}
\begin{eqnarray}
\lefteqn{
 e_1^3 e_3 - 2 e_1 e_2 e_3 - e_1^2 e_2 e_3 + e_2^2 e_3 + e_3^2 + e_1 e_3^2 + e_1^2 e_3^2 - e_2 e_3^2 
} \nonumber \\
& \hspace*{1.5in} = &
 q\left(e_2 - 3 e_3 + e_1 e_3 - e_2 e_3 + e_3^2\right).
\end{eqnarray}

It will also be useful to derive a $q$-independent combination of 
these expressions.  Let $P_a$ denote the difference
\begin{equation}
P_a \: = \: q x_a^4 \left( \prod_{c \neq a} \left( 1 - x_a^{-1} x_b^{-1} \right)
\right) 
\: - \: \left( 1 - x_a \right)^6,
\end{equation}
which vanishes from the equation of motion~(\ref{eq:lg36:phys}).
Then, consider the difference
\begin{equation}
x_3^2 \frac{ P_1 x_2^3 - P_2 x_1^3 }{ (x_1 x_2 - 1)(x_1 - x_2) }
\: - \:
x_2^2 \frac{ P_1 x_3^3 - P_3 x_1^3 }{ (x_1 x_3 - 1)(x_1 - x_3) }
\: = \: 0.
\end{equation}
Factoring out factors such as $(x_a - x_b)$ and $(1 - x_a x_b)$
(which are nonzero because of the excluded locus),
we are left with the $q$-independent constraint
\begin{equation}
e_2(x) - 6 e_3(x) + e_3(x) e_1(x) \: = \: 0,
\end{equation}
which can be written in terms of shifted Wilson line variables as
\begin{equation}  \label{eq:lg36:qind-phys}
e_1(z)^2 - 2 e_2(z) - e_1(z) e_2(z) + 3 e_3(z) + e_1(z) e_3(z)
\: = \: 0.
\end{equation}

\subsubsection{Comparison to $\lambda_y$ relations}
\label{sect:ex:lg36:lambda}

For $LG(3,6)$, we have the relations\footnote{
As observed previously, physics and mathematics conventions 
for
$LG(2,4)$ (but not $LG(3,6)$ or ordinary Grassmannians)
differ 
on
$q$.
}
\begin{equation}
\lambda_y(S) \star \lambda_y(Q) \: = \: 
(1+y)^6 \: - \: q
y^2 \left( 1 + y Q + y^2( 1 + \wedge^2 Q) + y^3 Q + y^4 \right),
\end{equation}
as given earlier in equation~(\ref{eq:lg:ex36intro}).
From the coefficients of powers of $y$, we have
\begin{eqnarray}
e_1(x) + e_1(\tilde{x}) & = & 6,   \label{eq:lg36:lambda:1}
\\
e_2(x) + e_1(x) e_1(\tilde{x}) + e_2(\tilde{x}) & = & 15-q,
\\
e_3(x) + e_2(x) e_1(\tilde{x}) + e_1(x) e_2(\tilde{x}) + e_3(\tilde{x}) & = &
20 - q e_1(\tilde{x}),
\\
e_3(x) e_1(\tilde{x}) + e_2(x) e_2(\tilde{x}) + e_1(x) e_3(\tilde{x})
& = & 15 - q\left( 1 + e_2(\tilde{x}) \right),
\\
e_3(x) e_2(\tilde{x}) + e_2(x) e_3(\tilde{x}) & = & 6 - q e_1(\tilde{x}),
\\
e_3(x) e_3(\tilde{x}) & = & 1-q,
\end{eqnarray}
where we have, formally, applied the splitting principle to write
$S = \oplus_a x_a$ and $Q = \oplus_{\ell} \tilde{x}_{\ell}$.

Using the first three relations to eliminate $e_{1,2,3}(\tilde{x})$,
and plugging back in, the remaining three equations become
\begin{eqnarray}
\lefteqn{
 20 e_1(x) - 15 e_1(x)^2 + 6 e_1(x)^3 - e_1(x)^4 + 15 e_2(x) - 12 e_1(x) e_2(x) + 3 e_1(x)^2 e_2(x)
} \nonumber \\
& & - e_2(x)^2 + 6 e_3(x) - 2 e_1(x) e_3(x)
\nonumber \\
& \hspace*{1.4in} = &
 15 - 16 q + 12 e_1(x) q - 3 e_1(x)^2 q + 2 e_2(x) q + q^2,
\label{eq:lg36:xr1}
\end{eqnarray}
\begin{eqnarray}
\lefteqn{
20 e_2(x) - 15 e_1(x) e_2(x) + 6 e_1(x)^2 e_2(x) - e_1(x)^3 e_2(x) - 6 e_2(x)^2 + 2 e_1(x) e_2(x)^2
} \nonumber \\
& &
+ 15 e_3(x) - 6 e_1(x) e_3(x) + e_1(x)^2 e_3(x) - 2 e_2(x) e_3(x)
\nonumber \\
& \hspace*{1.2in} = &
 6- 6 q + e_1(x) q + 6 e_2(x) q - 2 e_1(x) e_2(x) q + e_3(x) q,
\label{eq:lg36:xr2}
\end{eqnarray}
\begin{eqnarray}
\lefteqn{
 20 e_3(x) - 15 e_1(x) e_3(x) + 6 e_1(x)^2 e_3(x) - e_1(x)^3 e_3(x) - 6 e_2(x) e_3(x)
} \nonumber \\
& &
+ 2 e_1(x) e_2(x) e_3(x) - e_3(x)^2 
\nonumber \\
& \hspace*{1.5in} = &
 1- q + 6 e_3(x) q - 2 e_1(x) e_3(x) q.
\label{eq:lg36:xr3}
\end{eqnarray}

In addition, there is also a relation we can derive classically
from equation~(\ref{eq:lg36:lambda:1}).  Since $S^* \cong Q$,
we can take $\tilde{x}_{\ell} = x_a^{-1}$, hence
\begin{equation}
e_3(x) e_1(\tilde{x}) = e_2(x).
\end{equation}
Multiplying equation~(\ref{eq:lg36:lambda:1}) by $e_3(x)$, we then get
\begin{equation}  \label{eq:lg36:class1}
e_1(x) e_3(x) + e_2(x) \: = \: 6 e_3(x).
\end{equation}

Using the dictionary of section~\ref{sect:useful-identities},
we can rewrite the three relations~(\ref{eq:lg36:xr1})-(\ref{eq:lg36:xr3})
above in a basis of shifted Wilson
lines, as
\begin{eqnarray}
\lefteqn{
e_1(z)^4 - 3 e_1(z)^2 e_2(z) + e_2(z)^2 + 2 e_1(z) e_3(z)
} \nonumber \\
& \hspace*{0.5in} = &
q - 2 q e_1(z)  + 3 q e_1(z)^2  - 2 q e_2(z)  - q^2,
\end{eqnarray}
\begin{eqnarray}
\lefteqn{
 2 e_1(z)^4 - 6 e_1(z)^2 e_2(z) - e_1(z)^3 e_2(z) + 2 e_2(z)^2 + 2 e_1(z) e_2(z)^2
} \nonumber \\
& &
+ 4 e_1(z) e_3(z) + e_1(z)^2 e_3(z) - 2 e_2(z) e_3(z) 
\nonumber \\
& \hspace*{0.5in} = & 
2 q - 4 q e_1(z)  + 4 q e_1(z)^2  - q e_2(z) - 2 q e_1(z) e_2(z) + q e_3(z) ,
\end{eqnarray}
\begin{eqnarray}
\lefteqn{
e_1(z)^4 - 3 e_1(z)^2 e_2(z) - e_1(z)^3 e_2(z) + e_2(z)^2 + 2 e_1(z) e_2(z)^2 + 2 e_1(z) e_3(z)
} \nonumber \\
& &
 + e_1(z)^2 e_3(z) + e_1(z)^3 e_3(z) - 2 e_2(z) e_3(z) - 2 e_1(z) e_2(z) e_3(z) + e_3(z)^2
\nonumber \\
& \hspace*{0.5in} = &
q - 2 q e_1(z) + 2 q e_1(z)^2 - 2 q e_1(z) e_2(z) + 2 q e_1(z) e_3(z) ,
\label{eq:lg36:lambda:3}
\end{eqnarray}
and equation~(\ref{eq:lg36:class1}) becomes
\begin{equation}
 e_1(z)^2 - 2 e_2(z) - e_1(z) e_2(z) + 3 e_3(z) + e_1(z) e_3(z)
\: = \: 0.
\end{equation}
This last equation matches the physics prediction~(\ref{eq:lg36:qind-phys}).
Using this last equation, it is straightforward to see
the $\lambda_y$-class prediction~(\ref{eq:lg36:lambda:3}) above is equivalent to
the physics prediction~(\ref{eq:lg36:physpred:1}).

\section{Conclusions}

In this paper we have discussed various predictions for
quantum K theory from physics.  We first discussed some new bases
for the quantum K theory of ordinary Grassmannians, in terms of
shifted Wilson lines and $\lambda_y$ classes, which are naturally
related to physics computations.  We then turned to symplectic Grassmannians,
where we used physics to make propose descriptions for the quantum K theory of
symplectic Grassmannians (in terms of shifted Wilson lines)
and another basis for the quantum K theory of Lagrangian Grassmannians
(in terms of $\lambda_y$ classes), which we intend to 
study mathematically
in \cite{leom}.

\section{Acknowledgements}

We would like to thank M.~Bullimore, C.~Closset, and H.~Kim
for many useful conversations and 
collaboration on early versions of this paper.
We would also 
like to thank R.~Donagi, H.~Jockers, S.~Katz, and Y.-P.~Lee 
for useful discussions.
E.S. was partially supported by NSF grant PHY-1720321.

\appendix

\section{Tables of $LG(3,6)$ results}
\label{app:tables}

In this appendix we collect several pertinent facts
concerning the quantum K theory of $LG(3,6)$.

Classically, Wilson lines $W_T$ (Schur polynomials in $x$)
and Schubert cycles ${\cal O}_T$ are
related as follows:
\begin{eqnarray}
W_{\tiny\yng(1)} & = & 3 - {\cal O}_{\tiny\yng(1)} -
{\cal O}_{\tiny\yng(2)} - {\cal O}_{\tiny\yng(3)},
\\ 
W_{\tiny\yng(2)} & = & 6 - 4 {\cal O}_{\tiny\yng(1)} - 3 {\cal O}_{\tiny\yng(2)}
+ {\cal O}_{\tiny\yng(2,1)} - 3 {\cal O}_{\tiny\yng(3)} +
{\cal O}_{\tiny\yng(3,1)} + {\cal O}_{\tiny\yng(3,2)},
\\
W_{\tiny\yng(3)} & = & 10
- {\cal O}_{\tiny\yng(3,2,1)} - 6 {\cal O}_{\tiny\yng(3)} -
10 {\cal O}_{\tiny\yng(1)} - 5 {\cal O}_{\tiny\yng(2)} 
+ 5 {\cal O}_{\tiny\yng(2,1)} + 4 {\cal O}_{\tiny\yng(3,1)}
+ 3 {\cal O}_{\tiny\yng(3,2)},
\\
W_{\tiny\yng(2,1)} & = & 8 +
{\cal O}_{\tiny\yng(3)} - 8 {\cal O}_{\tiny\yng(1)} 
- 2 {\cal O}_{\tiny\yng(2)} + {\cal O}_{\tiny\yng(2,1)},
\\
W_{\tiny\yng(3,1)} & = & 15 +
3 {\cal O}_{\tiny\yng(3)} - 20 {\cal O}_{\tiny\yng(1)} 
+ 5 {\cal O}_{\tiny\yng(2,1)} - {\cal O}_{\tiny\yng(3,1)} -
{\cal O}_{\tiny\yng(3,2)},
\\
W_{\tiny\yng(3,2)} & = & 15 +
{\cal O}_{\tiny\yng(3,2,1)} + 12 {\cal O}_{\tiny\yng(3)}
- 25 {\cal O}_{\tiny\yng(1)} + 10 {\cal O}_{\tiny\yng(2)}
- 10 {\cal O}_{\tiny\yng(3,1)} - 3 {\cal O}_{\tiny\yng(3,2)},
\\
W_{\tiny\yng(3,2,1)} & = & 8
-3 {\cal O}_{\tiny\yng(3)} - 16 {\cal O}_{\tiny\yng(1)} + 
14 {\cal O}_{\tiny\yng(2)} 
- 5 {\cal O}_{\tiny\yng(2,1)} + {\cal O}_{\tiny\yng(3,1)}
+ {\cal O}_{\tiny\yng(3,2)},
\end{eqnarray}
and shifted Wilson lines $SW_T$ (Schur polynomials in $1-x$)
are related to Schubert cycles ${\cal O}_T$ classically as follows:
\begin{eqnarray}
SW_{\tiny\yng(1)} & = & 
{\cal O}_{\tiny\yng(1)} + {\cal O}_{\tiny\yng(2)} +
{\cal O}_{\tiny\yng(3)},
\\
SW_{\tiny\yng(2)} & = &
{\cal O}_{\tiny\yng(3)} + {\cal O}_{\tiny\yng(2)} +
{\cal O}_{\tiny\yng(2,1)} + {\cal O}_{\tiny\yng(3,1)} +
{\cal O}_{\tiny\yng(3,2)},
\\
SW_{\tiny\yng(3)} & = &
{\cal O}_{\tiny\yng(3)} + {\cal O}_{\tiny\yng(3,1)} +
2 {\cal O}_{\tiny\yng(3,2)} + {\cal O}_{\tiny\yng(3,2,1)},
\\
SW_{\tiny\yng(2,1)} & = &
{\cal O}_{\tiny\yng(3)} + {\cal O}_{\tiny\yng(2,1)} +
2 {\cal O}_{\tiny\yng(3,1)} + 2 {\cal O}_{\tiny\yng(3,2)},
\\
SW_{\tiny\yng(3,1)} & = &
{\cal O}_{\tiny\yng(3,1)} + 3 {\cal O}_{\tiny\yng(3,2)}
+ 2 {\cal O}_{\tiny\yng(3,2,1)},
\\
SW_{\tiny\yng(3,2)} & = &
{\cal O}_{\tiny\yng(3,2)} + 2 {\cal O}_{\tiny\yng(3,2,1)},
\\
SW_{\tiny\yng(3,2,1)} & = &
{\cal O}_{\tiny\yng(3,2,1)}.
\end{eqnarray}
This relations can be inverted as
\begin{eqnarray}
 {\cal O}_1 & = & SW_1 - SW_2 + SW_{2,1} - SW_3 + SW_{3,2} - SW_{3,2,1}, \\     
 {\cal O}_2 & = & SW_2 - SW_{2,1} + SW_{3,1} - 2 SW_{3,2} + 2 SW_{3,2,1}, \\
 {\cal O}_3 & = & SW_3 - SW_{3,1} + SW_{3,2} - SW_{3,2,1}, \\
 {\cal O}_{2,1} & = & SW_{2,1} - SW_3 - SW_{3,1} + 3 SW_{3,2} - 3 SW_{3,2,1}, \\
 {\cal O}_{3,1} & = & SW_{3,1} - 3 SW_{3,2} + 4 SW_{3,2,1}, \\
 {\cal O}_{3,2} & = & SW_{3,2} - 2 SW_{3,2,1}, \\
 {\cal O}_{3,2,1} & = & SW_{3,2,1},
\end{eqnarray}
where for notational brevity we have indicated Young tableau by the number
of boxes in each row.

The (quantum-corrected) products of Schubert classes arising in mathematics
are given by
\begin{align}
        & {{\cal O}_{1}}^2  =  -{\cal O}_{3}+2 {\cal O}_2 + {\cal O}_{3,1} - {\cal O}_{2,1},
 && {\cal O}_{1} \cdot {\cal O}_{2} = -q+q{\cal O}_{1}+2{\cal O}_{3}+ {\cal O}_{2,1} - 2{\cal O}_{3,1},
\\
        & {\cal O}_{1} \cdot {\cal O}_{3}  =  q - q{\cal O}_{1} + {\cal O}_{3,1},
&& {\cal O}_{1} \cdot {\cal O}_{2,1} = -q{\cal O}_{1}+ q{\cal O}_{2}+2{\cal O}_{3,1} - {\cal O}_{3,2},
\\
        & {\cal O}_{1}\cdot {\cal O}_{3,1}  =  q{\cal O}_{1} - 2 q {\cal O}_{2} + q {\cal O}_{2,1} 
&& {\cal O}_{1}\cdot {\cal O}_{3,2} =q {\cal O}_{2} - q {\cal O}_{2,1} + {\cal O}_{3,2,1},
\\
        & \hspace*{0.8in} +2{\cal O}_{3,2} - {\cal O}_{3,2,1}, \nonumber
\\
        & {\cal O}_{1}\cdot {\cal O}_{3,2,1}  =  q{\cal O}_{2,1},
&& {{\cal O}_{2}}^2 = q -2q{\cal O}_{1}+q{\cal O}_{2}+2{\cal O}_{3,1}-{\cal O}_{3,2},
\\
        & {\cal O}_{2} \cdot {\cal O}_{3}  =  q{\cal O}_{1}-q{\cal O}_{2} + {\cal O}_{3,2},
&& {\cal O}_{2}\cdot {\cal O}_{2,1} = q{\cal O}_{1} -2q{\cal O}_{2} + q{\cal O}_{2,1} + 2{\cal O}_{3,2} - {\cal O}_{3,2,1},
\\
        & {\cal O}_{2}\cdot {\cal O}_{3,1}  =  2q {\cal O}_{2} -q{\cal O}_{3} -2q {\cal O}_{2,1} 
&& {\cal O}_{2}\cdot {\cal O}_{3,2} = q{\cal O}_{3} + q{\cal O}_{2,1} - q{\cal O}_{3,1},
\\
        & \hspace*{0.8in} + q{\cal O}_{3,1}+{\cal O}_{3,2,1}, \nonumber
\\
        & {\cal O}_{2}\cdot {\cal O}_{3,2,1}   =  q {\cal O}_{3,1},
&& {{\cal O}_{2,1}}^2 = 2q {\cal O}_{2}-q {\cal O}_{3}-q {\cal O}_{2,1} + q{\cal O}_{3,1},
\\
        & {\cal O}_{2,1} \cdot {\cal O}_{3,1}  =  -q^2 + q^2 {\cal O}_{1}+2q {\cal O}_{3}
&& {\cal O}_{2,1}\cdot {\cal O}_{3,2} = q^2-q^2 {\cal O}_{1}+q {\cal O}_{3,1},
\\
        & \hspace*{0.8in} +q {\cal O}_{2,1}-2q {\cal O}_{3,1}, \nonumber
\\
        & {\cal O}_{2,1}\cdot {\cal O}_{3,2,1}  =  q^2 {\cal O}_{1},
&& {{\cal O}_{3}}^2  = q{\cal O}_{2},
\\
        & {\cal O}_3 \cdot {\cal O}_{2,1} = q {\cal O}_2 - q {\cal O}_{2,1} + {\cal O}_{3,2,1},
\\
        & {\cal O}_{3}\cdot {\cal O}_{3,1}  =  q {\cal O}_{3} + q {\cal O}_{2,1}-q {\cal O}_{3,1},
&& {\cal O}_{3} \cdot {\cal O}_{3,2} = q {\cal O}_{3,1},
\\
        & {\cal O}_{3}\cdot {\cal O}_{3,2,1}   =  q {\cal O}_{3,2},
&& {{\cal O}_{3,1}}^2 = q^2-2q^2 {\cal O}_{1}+q^2 {\cal O}_{2}+2q{\cal O}_{3,1}-q{\cal O}_{3,2},
\\
        & {\cal O}_{3,1}\cdot {\cal O}_{3,2}   =  q^2 {\cal O}_{1}-q^2 {\cal O}_{2}+q{\cal O}_{3,2},
&& {\cal O}_{3,1}\cdot {\cal O}_{3,2,1} = q^2 {\cal O}_{2},
\\
        & {{\cal O}_{3,2}}^2  =  q^2 {\cal O}_{2},
&& {\cal O}_{3,2}\cdot {\cal O}_{3,2,1} = q^2{\cal O}_{3},
\\
        & {{\cal O}_{3,2,1}}^2  =  q^3.
\end{align}

\end{document}